\documentclass[a4paper]{article}
\usepackage{a4wide}
\usepackage{graphicx}
\graphicspath{{./}}
\usepackage{amsthm}
\usepackage{amssymb, latexsym, mathrsfs}
\usepackage{amsmath}
\usepackage{dsfont}
\usepackage{enumerate}
\usepackage{subcaption}
\usepackage{algorithm}
\usepackage{color}
\usepackage{url}
\usepackage[affil-it]{authblk}
\usepackage{booktabs}
\theoremstyle{plain}

\newtheorem{rmk}{Remark}

\newtheorem{lem}{Lemma}

\usepackage{tikz}
\usepackage{schemabloc}

\newcommand{\Eset}{\mathbb{E}}

\newcommand{\Iset}{\mathbb{I}}

\newcommand{\Rset}{\mathbb{R}}

\newcommand{\hc}{{\hat{c}}}

\newcommand{\hu}{{\hat{u}}}

\newcommand{\hB}{{\hat{B}}}

\newcommand{\hD}{{\hat{D}}}

\newcommand{\hR}{{\hat{R}}}

\newcommand{\bv}{{\bar{v}}}

\newcommand{\bx}{{\bar{x}}}
\newcommand{\by}{{\bar{y}}}

\newcommand{\tp}{{\tilde{p}}}

\newcommand{\tv}{{\tilde{v}}}

\newcommand{\tJ}{{\tilde{J}}}

\newcommand{\tV}{{\tilde{V}}}

\newcommand{\AAA}{{\mathcal{A}}}
\newcommand{\BB}{{\mathcal{B}}}

\newcommand{\GG}{{\mathcal{G}}}

\newcommand{\NN}{{\mathcal{N}}}

\newcommand{\PP}{{\mathcal{P}}}
\newcommand{\QQ}{{\mathcal{Q}}}
\newcommand{\RR}{{\mathcal{R}}}

\newcommand{\diag}{{\mbox{diag}}}                  
\newcommand{\abs}[1]{{|{#1}|}}                     
\newcommand{\norme}[2]{||{#1}||_{#2}}            

\newcommand{\mbf}[1]{\mathbf{#1}}                  
\newcommand{\One}{\textbf{1}}

\newcommand{\matr}[1]{
\begin{bmatrix}
    #1
\end{bmatrix}
}

\begin{document}

	\title{Model predictive controllers for reduction of mechanical fatigue in wind farms}

    \author[2]{Stefano Riverso%
       \thanks{Electronic address: \texttt{riverss@utrc.utc.com}}} 

	\author[1]{Simone Mancini%
       \thanks{Electronic address: \texttt{simone.mancini01@universitadipavia.it}}}

     \author[1]{Fabio Sarzo%
       \thanks{Electronic address: \texttt{fabio.sarzo01@universitadipavia.it}}}

     \author[1]{Giancarlo Ferrari-Trecate%
       \thanks{Electronic address: \texttt{giancarlo.ferrari@unipv.it}; Corresponding author}}

     \affil[1]{Dipartimento di Ingegneria Industriale e dell'Informazione\\Universit\`a degli Studi di Pavia}
     \affil[2]{United Technologies Research Center Ireland}  
     \date{\textbf{Technical Report}\\ March, 2015}

     \maketitle

       \begin{abstract}
We consider the problem of dispatching WindFarm (WF) power demand to individual Wind Turbines (WT) with the goal of minimizing mechanical stresses. We assume wind is strong enough to let each WTs to produce the required power and propose different closed-loop Model Predictive Control (MPC) dispatching algorithms. Similarly to existing approaches based on MPC, our methods do not require changes in  WT hardware but only software changes in the SCADA system of the WF. However, differently from previous MPC schemes, we augment the model of  a WT with an ARMA predictor of the wind turbulence, which reduces uncertainty in wind predictions over the MPC control horizon. This allows us to develop both stochastic and deterministic MPC algorithms. In order to compare different MPC schemes and demonstrate improvements with respect to classic open-loop schedulers, we performed simulations using the SimWindFarm toolbox for MatLab.
We demonstrate that MPC controllers allow to achieve reduction of stresses even in the case of large installations such as the 100-WTs Thanet offshore WF.
	
	\emph{Key Words: } Wind farm control, Model predictive control, Stochastic control, ARMA models.
     \end{abstract}

     \newpage

     \section{Introduction}
          \label{sec:intro}
\begin{figure}[!htb]
            \centering
            \includegraphics[scale=0.4]{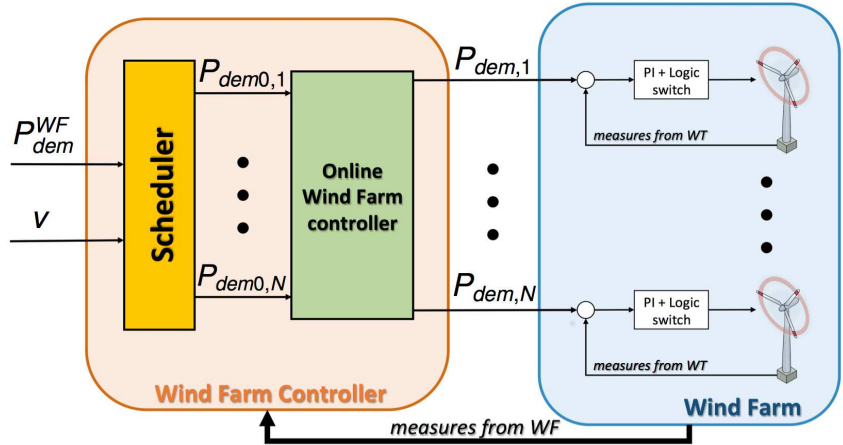}
            \caption{Wind farm controller. Inputs are the total power demand and, optionally, wind measurements.}
            \label{fig:WFcontr}
          \end{figure}

          In the last few years, the interest in wind energy has been constantly increasing. From the end $2010$ to mid-$2013$ the global wind capacity grew up by $48.3\%$, generating around $3.5\%$ of the world electricity demand \cite{WorldWindEnergyAssociation2013}.  It has been estimated that, at the end of $2013$, the worldwide wind capacity has reached $318\;[GW]$. 
This constant increase of Wind Farms (WFs) installations is due to the fact that wind energy is an excellent environmental and friendly solution to the problem of energy shortage. For example, three years after the nuclear disaster of Fukushima, local Japan government, in particular the Fukushima prefecture, is considering to supply their regions with $100\%$ renewable energy by $2040$ \cite{JapReg}. To achieve this goal, despite the increase of the installed capacity of wind turbines, it is necessary to face new engineering and science challenges to improve efficiency and durability of the systems. 
In order to maximize the economic investment and the power generation efficiency, the size of WTs will be increased so as to produce more than $20\;[MW]$. These larger WTs will be installed both in onshore and offshore environments, subjecting their flexible structures to forces of different entities. To face these problems, we need advanced control architectures: the aim is to improve the efficiency by reducing structural stress and hence extending the lifetime of components. 
Indeed, ``the lifetime of wind power plants is considered to be about 30 years, even if usually after 20 years these plants are dismantled because of the progressive decrease in the energy production due to the aging of wind turbine components'' \cite{ABB2011}.\\
          WF control is essential to fit the required power, maximizing the performance, minimizing the mechanical forces acting on WTs and to detect anomalies in the WTs \cite{Pao2011} and \cite{Odgaard2013}. The required power $P_{dem}^{WF}$ is determined by a network operator, who specifies this value as a function of national load profiles and other economic and political criteria. However, the power that can be actually produced by a WF strictly depends on the wind blowing on its WTs. In this respect, each WT can work in two different operating regions. The first one is called power maximization region and it is selected when the wind acting on the WTs is not strong enough to ensure the production of the required power. The second one, called power tracking region, is selected if the wind blowing on the WT is enough to produce the demanded power. 
When all WTs are in the power tracking region, the use of a WF controller can bring major advantages, as one can choose different strategies to dispatch the power demand among WTs. A first simple solution is represented by the adoption of a scheduler: given the wind speed profile that acts on the WF and the power demand provided by the network operator, the scheduler divides the power demand according to an open-loop strategy, e.g. distributing the power equally between the WT or, alternatively, activating the smallest number of WTs. However, this could lead some WTs to be stressed much more than others.
For these reasons, it is 
convenient to introduce a closed-loop WF controller, that uses the on-line measurements from the WTs (see Figure \ref{fig:WFcontr}).
          
          In literature, there exist different approaches to the design of WF controllers. A first idea, explained in detail in \cite{Grunnet2010}, is to use the knowledge of the available power $P_{a}$ for each WT dispatching the $P_{dem}^{WF}$ proportionally to $P_{a}$. However, this approach could increase the tower oscillations and the stresses imposed on the motor shaft. In \cite{Spudic2011}, the authors propose a linearized WT model and a Model Predictive Control (MPC) scheme that evaluates, on a given prediction horizon, the power demand set-points to achieve different aims such as mechanical stress reduction. An advantage of this controller is the possibility to force the fulfillment of input constraints. In this case, the stochasticity introduced by the wind is not considered as part of the model, but it is assumed that the wind profile is known a priori, leading to a deterministic approach. 
From \cite{Burton2001}, we can assume that the wind is the sum of an average speed, which change in the order of hours, and of a zero mean turbulence variation, that changes on a faster time scale. Using this assumption, low-pass filters are introduced in \cite{Madjidian2011,Biegel2011,Biegel2013a} in order to model the wind turbulence. In these papers, the authors propose stochastic WF controllers. However, they do not consider constraints on the references of power demand due to the mechanical characteristics of the WT and the instantaneous variations of the wind. It is worth noting that all the predictive control techniques discussed above allow to reduce the fatigue on the WF by making software changes on the SCADA systems and they do not require to replace hardware component of WTs. More recently, same authors of \cite{Spudic2011} proposed a supervisory controller which can be easily installed on very large WFs \cite{Spudic2015}. For sake of completeness, it should be noted that in the literature there are examples of control schemes implemented directly on the WT (see for example \cite{Gros2013a} and \cite{Betti2014}), but these approaches require substantial investments to upgrade existing WFs.

          In this paper, we propose new WF controllers using MPC regulators. As a reference model for a WT in the power tracking region we use a linearized version of the NREL model \cite{Jonkman2009}. 
Differently from the previous approaches, we will account for the wind variations assuming  a Kaimal wind turbulence spectrum \cite{Burton2001} and modeling turbulence as an ARMA process. Then, we propose an optimal one-step-ahead predictor computed from the ARMA process. This allows us to develop Deterministic MPC (DMPC) and Stochastic MPC (SMPC) regulators with the goal of dispatching power demands between WTs so as to minimize tower bending and fatigue on the motor shaft while guaranteeing that the sum of the power demand for each WT meets $P_{dem}^{WF}$.
In particular, we will use the SMPC scheme proposed in \cite{Magni2009a} and \cite{Farina2013b} in order to account for wind stochasticity and we also design two different DMPC regulators which will not account for the variance of wind turbulence. In this paper we will not make use of experimental data and, hence, to evaluate performance of the proposed MPC controllers, we perform several simulations using MatLab/Simulink and the SimWindFarm (SWF) toolbox \cite{Grunnet2010}. 

          The paper is organized as follows. In Section \ref{sec:windFarmModel} we introduces the WF model, by proposing a linearized model of the adopted WT and an optimal one-step-ahead predictor for wind turbulence. In Section \ref{sec:mpcpre} we propose a SMPC regulator and two DMPC regulators. In Section \ref{sec:simresults} we present simulation results and Section \ref{sec:conclusions} is dedicated to some conclusions and possible future improvements.

          \textbf{Notation.} We use $a:b$ for the set of integers $\{a,a+1,\ldots,b\}$. The column vector with $s$ components $v_1,\dots,v_s$ is $\mbf v=(v_1,\dots,v_s)$. The function $\diag(G_1,\ldots,G_s)$ denotes the block-diagonal matrix composed by $s$ block $G_i$, $i=1,\ldots,s$. Moreover, $tr(Q)$ is the trace of matrix $Q$. The symbol $\One_r$ denotes a column vector in $\Rset^r$ with all elements equal to $1$. Furthermore, $\Iset$ is the identity matrix. We use $\norme{x}{P}$ to define the $P$-weighted seminorm, defined $\mbox{for all }x\in\Rset^n$ by $\norme{x}{P}=x^TPx$, where $P$ is a positive-semidefinite real symmetric matrix. The functions $\Eset[\cdot]$, $var[\cdot]$ and $cov[\cdot]$ denote mean value, variance and covariance of random variables. The function $std[\mbf v]$, where $\mbf v=(v_1,\dots,v_s)$, denotes the sample standard deviation of measurements $v_i,~i=1:s$. The function $\PP(A)$ denotes the probability of the event $A$. The function $WGN(\alpha,\beta)$ denotes White Gaussian Noise (WGN) with mean $\alpha$ and variance $\beta$. The standard normal distribution with mean $\alpha$ and variance $\beta$ is denoted with $\NN(\alpha,\beta)$.

     \section{WF model}
          \label{sec:windFarmModel}
          In this section, we propose a linearized model for the WF. We first introduce a linearized model of a WT operating in the tracking region and then we design an optimal predictor for the wind turbulence.

          \subsection{NREL WT model}
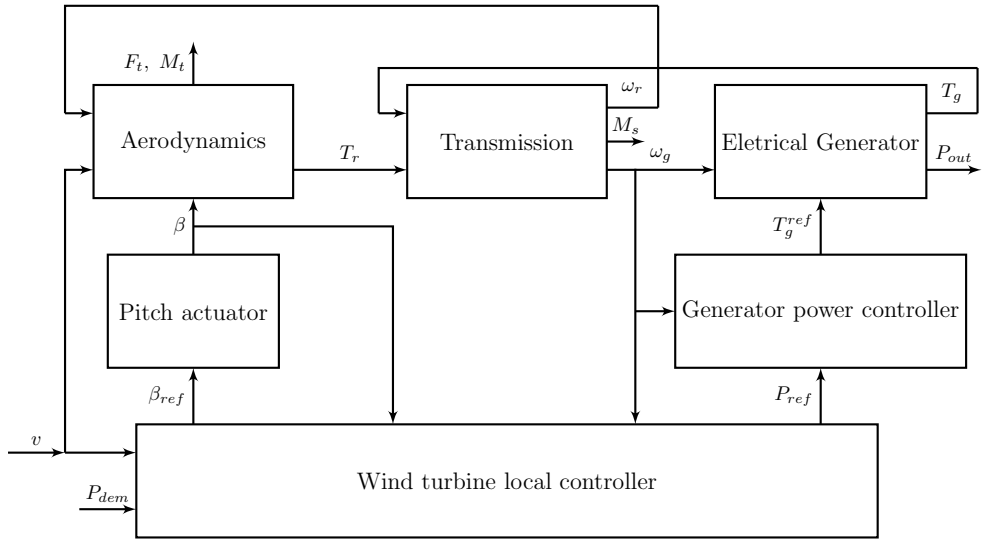
\begin{figure}[!htb]
                 \centering
                 \tikzstyle{input} = [coordinate]
\tikzstyle{output} = [coordinate]
\tikzstyle{guide} = []
\tikzstyle{block} = [draw, rectangle, minimum height=1cm]

\begin{tikzpicture}[thick,scale=0.75, every node/.style={transform shape}]
  \node [block,minimum height=2cm, minimum width=3.5cm] (aerodynamics) {\large{Aerodynamics}};
  \node [output, right of=aerodynamics,xshift=0.75cm,yshift=-0.5cm] (Tr_aer) {};
  \node [input, left of=aerodynamics,xshift=-0.75cm,yshift=0.5cm] (wr_aer) {};
  \node [input, left of=aerodynamics,xshift=-1.25cm,yshift=0.5cm] (wr_aer_pre) {};
  \node [input, left of=aerodynamics,xshift=-0.75cm,yshift=-0.5cm] (v_aer) {};
  \node [input, below of=aerodynamics] (beta_aer) {};
  \node [output, above of=aerodynamics,yshift=0cm] (Ft_aer) {};
  \node [guide, above of=Ft_aer,yshift=-0.1cm] (Ft_aer_end) {};

  \node [block,node distance=5.5cm,right of=aerodynamics,minimum height=2cm, minimum width=3.5cm] (transmission) {\large{Transmission}};
  \node [input, left of=transmission,xshift=-0.75cm,yshift=-0.5cm] (Tr_trans) {};
  \node [input, left of=transmission,xshift=-0.75cm,yshift=0.5cm] (Tg_trans) {};
  \node [input, left of=transmission,xshift=-1.25cm,yshift=0.5cm] (Tg_trans_pre) {};
  \node [output, right of=transmission,xshift=0.75cm,yshift=0.6cm](wr_trans) {};
  \node [output, right of=wr_trans,xshift=-0.1cm](wr_trans_end) {};
  \node [output, right of=wr_trans,xshift=-0.1cm,yshift=1.8cm] (wr_trans_end_up) {};    
  \node [output, right of=transmission,xshift=0.75cm,yshift=-0.5cm] (wg_trans) {};
  \node [guide, right of=wg_trans,xshift=-0.5cm,yshift=0.15cm] (wg_trans_right) {};
  \node [output, right of=transmission,xshift=0.75cm] (Ms_trans) {};
  \node [guide, right of=Ms_trans,xshift=-0.2cm] (Ms_trans_end) {};

  \node [block,node distance=5.5cm,right of=transmission,minimum height=2cm, minimum width=3.5cm] (eleGenerator) {\large{Eletrical Generator}};
  \node [input, left of=eleGenerator,xshift=-0.85cm,yshift=-0.5cm] (wg_gen) {};
  \node [input, below of=eleGenerator] (Tgref_gen) {};
  \node [output, right of=eleGenerator,xshift=0.85cm,yshift=-0.5cm] (pout_gen) {};
  \node [guide, right of=pout_gen,xshift=0.1cm] (pout_gen_end) {};
  \node [output, right of=eleGenerator,xshift=0.85cm,yshift=0.5cm] (Tg_gen) {};
  \node [output, right of=Tg_gen,xshift=-0.1cm] (Tg_gen_end) {};
  \node [output, right of=Tg_gen,xshift=-0.1cm,yshift=0.8cm] (Tg_gen_end_up) {};
  
  \node [block,node distance=3cm,below of=aerodynamics,minimum height=2cm, minimum width=3cm] (pitch) {\large{Pitch actuator}};
  \node [input, below of=pitch] (betaref_pitch) {};
  \node [output, above of=pitch] (beta_pitch) {};
  \node [output, above of=pitch,yshift=0.5cm] (beta_pitch_up) {};

  \node [block,node distance=3cm,below of=eleGenerator,minimum height=2cm, minimum width=3cm] (ctrlGen) {\large{Generator power controller}};
  \node [input, below of=ctrlGen] (Pref_ctrlGen) {};
  \node [output, above of=ctrlGen] (Tgref_ctrlGen) {};
  \node [input, left of=ctrlGen,xshift=-1.55cm] (wg_ctrlGen) {};
  \node [input, left of=ctrlGen,xshift=-2.25cm] (wg_ctrlGen_left) {};

  \node [block,node distance=6cm,below of=transmission,minimum height=2cm, minimum width=13cm] (ctrlWT) {\large{Wind turbine local controller}};
  \node [input, left of=ctrlWT,xshift=-5.5cm,yshift=0.5cm] (v_ctrlWT) {};
  \node [input, left of=ctrlWT,xshift=-6.75cm,yshift=0.5cm] (v_ctrlWT_left) {};
  \node [input, left of=ctrlWT,xshift=-7.75cm,yshift=0.5cm] (v_ctrlWT_left_left) {};
  \node [input, above of=ctrlWT, xshift=-2cm] (beta_ctrlWT) {};
  \node [input, above of=ctrlWT,xshift=2.25cm] (wg_ctrlWT) {};
  \node [output, above of=ctrlWT, xshift=5.5cm] (Pref_ctrlWT) {};
  \node [output, above of=ctrlWT, xshift=-5.5cm] (betaref_ctrlWT) {};
  \node [input, left of=ctrlWT,xshift=-5.5cm,yshift=-0.5cm](Pdem_ctrlWT) {};
  \node [input, left of=ctrlWT,xshift=-6.5cm,yshift=-0.5cm](Pdem_ctrlWT_left) {};

  \sbRelier[$T_r$]{Tr_aer}{Tr_trans}

  \sbRelier[$F_t,~M_t$]{Ft_aer}{Ft_aer_end}

  \sbRelier[$\beta_{ref}$]{betaref_ctrlWT}{betaref_pitch}
  \sbRelier[$\beta$]{beta_pitch}{beta_aer}
  \draw [draw,->,>=latex'] (beta_pitch_up) -| (beta_ctrlWT);

  \sbRelier[$P_{ref}$]{Pref_ctrlWT}{Pref_ctrlGen}
  \sbRelier[$T_g^{ref}$]{Tgref_ctrlGen}{Tgref_gen}
  \sbRelier[]{wg_ctrlGen_left}{wg_ctrlGen}
  
  \sbRelier[$P_{out}$]{pout_gen}{pout_gen_end}
  \draw [draw] (Tg_gen) -- node[yshift=0.35cm]{$T_g$} (Tg_gen_end);
  \draw [draw] (Tg_gen_end) -- (Tg_gen_end_up);
  \draw [draw] (Tg_gen_end_up) -| (Tg_trans_pre);
  \sbRelier[]{Tg_trans_pre}{Tg_trans}

  \sbRelier[$\omega_g$]{wg_trans}{wg_gen}
  
  \draw [draw,->,>=latex'] (wg_trans_right) -> (wg_ctrlWT);
  
  \sbRelier[$M_s$]{Ms_trans}{Ms_trans_end}

  \sbRelier[]{v_ctrlWT_left}{v_ctrlWT}
  \sbRelier[$v$]{v_ctrlWT_left_left}{v_ctrlWT_left}
  \draw [draw,->,>=latex'] (v_ctrlWT_left) |- (v_aer);

  \draw [draw] (wr_trans) -- node[yshift=0.35cm]{$\omega_r$} (wr_trans_end);
  \draw [draw] (wr_trans_end) -- (wr_trans_end_up);
  \draw [draw] (wr_trans_end_up) -| (wr_aer_pre);
  \sbRelier[]{wr_aer_pre}{wr_aer}

  \sbRelier[$P_{dem}$]{Pdem_ctrlWT_left}{Pdem_ctrlWT}

\end{tikzpicture}
                 \caption{Overview of the blocks of the NREL WT model.}
                 \label{fig:modelOvw}
               \end{figure}

               The NREL WT model is an offshore $5$-MW baseline variable speed wind turbine equipped with an active hydraulic pitch control. This model has been proposed in order to become a standard for large WTs. In this section, we derive a linearized model of the nonlinear system described in Figure \ref{fig:modelOvw}. In particular, we use the results described in \cite{Spudic2010a}, taking advantage of simplifications introduced in \cite{Grunnet2010}. We defer the interested reader to \cite{Jonkman2009} and \cite{Spudic2010a} for a complete description of the model. In the following sections, we describe each block of Figure \ref{fig:modelOvw}. Moreover, a variable with index $0$ means steady-state variable.

			\subsubsection{Aerodynamics}         
			   \label{subsub:aereo}      
               The conversion of the wind energy in available green energy can be described by an aerodynamic model of the rotor, hence we can model how the energy captured by the rotor can be converted into driving torque of the rotating machine. In the NREL WT model this transformation is represented by the following static nonlinear equation
               \begin{equation}
                 \label{eq:windTurbPow}
                 P_{a}(t) = \frac{\pi}{2} \rho R^{2} v(t)^{3} C_{P}\left(\lambda(t),\beta(t)\right),\qquad\text{with }\lambda(t) = \frac{\omega_{r}(t) R}{v(t)}
               \end{equation}
               where $P_{a}(t)$ is the wind turbine power [$W$], $\rho$ is the air density [$\frac{kg}{m^{3}}$], $R$ is the radius of wind turbine rotor [$m$], $v(t)$ is the wind speed [$\frac{m}{s}$], $C_{P}$ is the power coefficient, $\lambda(t)$ is the tip speed ratio, $\beta(t)$ is the collective pitch angle [$^{\circ}$], $\omega_{r}(t)$ is the rotational speed of wind turbine's rotor [$\frac{rad}{s}$]. The $C_{P}$ parameter in \eqref{eq:windTurbPow} is a characteristic nonlinear function depending on tip speed ratio and pitch angle of the blade. Furthermore, defining the aerodynamic torque applied to the rotor shaft $T_{r}(t) = \frac{P_{a}(t)}{\omega_{r}(t)}$ and replacing $P_{a}(t)$ using \eqref{eq:windTurbPow}, we obtain
               \begin{equation}
                 \label{eq:windTurRotShaft}
                 T_{r}(t) = \frac{\pi}{2} \rho R^{3} v(t)^{2} C_{Q}(\lambda(t),\beta(t)),\qquad C_{Q}(\lambda(t),\beta(t)) = \frac{C_{P}(\lambda(t),\beta(t))}{\lambda(t)}
               \end{equation}
               where $C_Q$ is the torque coefficient. During the conversion process, part of wind energy is dissipated through a secondary effect that acts on the rotor of the wind turbine. This force operates perpendicularly to respect to the rotor plane, producing a tower bending moment and, consequently, oscillations on the wind turbine. The force exerted is called thrust force ($F_{t}(t)$) and is modeled by the following nonlinear static relation
               \begin{equation}
                 \label{eq:windTurThrustForce}
                 F_{t}(t) = \frac{\pi}{2} \rho R^{2} v(t)^{2} C_{T}(\lambda(t),\beta(t)),\qquad M_t(t)=hF_t(t)
               \end{equation}
               where $C_{T}$ represent the thrust coefficient, $h$ is the tower height, $M_t(t)$ is the tower bending moment caused mainly by the thrust force $F_t(t)$, hence we do not consider any elastic force which could increase $M_t(t)$. \\
               From \eqref{eq:windTurRotShaft} and \eqref{eq:windTurThrustForce}, by linearization about the operating point  $v_{0}$, $\beta_{0}$, $\omega_{r0}$, $T_{r0}$, $M_{t0}$, we obtain the following linear models
               \begin{equation}
                 \label{eq:RotTorqLTI}
                 T_{r}(t) - T_{r0} = K_{v T_{r}}(v(t)-v_{0}) + K_{w T_{r}}(\omega_{r}(t)-\omega_{r0})+ K_{\beta T_{r}}(\beta(t)-\beta_{0})
               \end{equation}
               \begin{equation}
                 \label{eq:Mts}
                 M_{t}(t) - M_{t0} = K_{v M_{t}}(v(t)-v_{0}) + K_{\omega M_{t}}(\omega_{r}(t)-\omega_{r0}) + K_{\beta M_{t}} (\beta(t)-\beta_{0}).
               \end{equation}

			\subsubsection{Wind turbine local controller}  
               Each WT is equipped with a local controller. The NREL WT local control system is simpler than other WT controllers: indeed, it does not use wind speed measurements and, moreover, does not provide additional blocks for oscillation damping (see for example \cite{Pao2011} and references therein). The NREL WT control scheme consists of two tracking loops: the first to compute the power reference $P_{ref}(t)$ and the second to compute the pitch angle reference $\beta_{ref}(t)$, based on the measure of the rotational speed $\omega_{g}(t)$ of the generator. The NREL WT controller operates in the following configurations.
               \begin{itemize}
               \item\emph{Power tracking}. $P_{ref}(t)$, boosted to compensate the generator efficiency and constraints on the generator rated power, tracks $P_{dem}(t)$. $\beta_{ref}(t)$ is set by a PI regulator where the error is computed as the difference of $\omega_{g}(t)$ to respect to the steady-state rotational speed of the generator $\omega_{g0}$. The nominal gains of the PI are adapted online based on $P_{dem}(t)$ and $\beta_{ref}(t)$.
               \item\emph{Power maximization}. $\beta_{ref}(t)$ is fixed to zero and $P_{ref}(t)$ is evaluated through a nonlinear function implemented in a look-up table.
               \end{itemize}
               Furthermore, a switching logic alternates this two configurations under specific conditions. Since our aim is to control the set-point $P_{dem}(t)$ for each WT, in the following we will consider the power tracking configuration. In order to obtain a linearized model of the NREL WT, we need to study the static behavior of the WT. Since the WT operates in two different configurations, the static behavior is completely different. A detailed analysis is given in Section 4.2 of \cite{Spudic2010a}.\\
               Linearizing the local adaptive PI controller described above around the operating point $(P_{ref0},\beta_{ref0},\omega_{g0})$, we obtain the following linear model of the PI regulator for computing $\beta_{ref}(t)$
               \begin{align}
               		\label{eq:omegagfiltLTI}\frac{d}{dt}(\omega_{g}^{f}(t)-\omega_{g0})&=-\frac{1}{T_\omega}(\omega_{g}^{f}(t)-\omega_{g0})+\frac{1}{T_\omega}(\omega_{g}(t)-\omega_{g0})\\
               	 	\label{eq:betareomegagLTI}\frac{d}{dt}(\beta_{ref}(t)-\beta_{ref0})&=\frac{K_P-K_IT_\omega}{T_w}(\omega_{g}^{filt}(t)-\omega_{g0})-\frac{K_P}{T_\omega}(\omega_{g}(t)-\omega_{g0})	 
               \end{align}
               where $T_{\omega}$ is a time constant of first order low-pass filter, that lumps the effects of the measurement device, $\omega_{g}^{f}$ is the filtered rotational speed of the generator and $K_P$ and $K_I$ are the gains of the PI controller corresponding to $P_{ref0}$ and $\beta_{ref0}$. Note that, asymptotically, $\omega_{g}^{f}=\omega_{g}=\omega_{g0}$.
               
			\subsubsection{Transmission}                
                The transmission system is a MIMO linear system describing the stiffness and damping of the low speed shaft generator, hence describing, through damped harmonic oscillators, the effects of $T_r$ and $T_g$ (generator torque) on $\omega_r$, $\omega_g$ and $M_s$ (main shaft torque). This part of the system can be modeled as a shaft with lumped inertia, omitting the fast dynamics related to the shaft elasticity. Therefore, using results in \cite{VanderHooft2003}, the following linear relations are obtained
               \begin{align}
                 \label{eq:wrLTIpart}\frac{d}{dt}(\omega_{r}(t)-\omega_{r0})&=\frac{1}{J_{r}+n_{gb}^{2}J_{g}}\left((T_{r}(t)-T_{r0}) - n_{gb}(T_{g}(t)-T_{g0})\right) \\
                 \label{eq:wgLTIpart}\omega_{g}(t)&=n_{gb} w_{r}(t)\\
                 (M_{s}(t)-M_{s0}) &= \frac{n_{gb}J_{r}}{J_{r}+n_{gb}^{2}J_{g}}(T_{g}(t)-T_{g0}) + \frac{n_{gb}^{2}}{J_{r}+n_{gb}^{2}J_{g}}(T_{r}(t)-T_{r0})\label{eq:MsLTIpart}
               \end{align}
               where $n_{gb}$ is the multiplication ratio of the gearbox and $J_g$ and $J_r$ are the inertia of generator and rotor, respectively.
               
            \subsubsection{Pitch actuator}
				The pitch actuator drives $\beta$ to $\beta_{ref}$. This variation is carried out via a servo drive that moves each blade on $\beta_{ref}$. This set-point is reached using hydraulic pitch actuator. However, for the design of WF controller, we can assume that $\beta\approx\beta_{ref}$.

            \subsubsection{Generator power controller}
			   In the NREL WT, the output electrical power $P(t)$ is modeled by the static nonlinear equation
               \begin{equation*}
                 \label{eq:NrelPow}
                 P(t) = \mu \omega_{g}(t) T_{g}(t)         
               \end{equation*}
               where $\mu$ is the generator efficiency. Therefore the generator power controller can easily compute the generator torque reference $T_g^{ref}$ as
               \begin{equation}
                 \label{eq:Tgref}
                 T_g^{ref}(t) = \frac{P_{ref}(t)}{\mu \omega_{g}(t)}.
               \end{equation}
               Moreover, a linearized model of equation \eqref{eq:Tgref} is
               \begin{equation}
                 \label{eq:Tglinear}
                 T_{g}^{ref}(t)-T_{g0} = \frac{1}{\mu \omega_{g0}}(P_{ref}(t)-P_{ref0}) - \frac{P_{ref0}}{\mu \omega_{g0}^2}(\omega_{g}(t)-\omega_{g0}).
               \end{equation}            
            
			\subsubsection{Electrical generator}            
               The generator dynamics is described by a lower pass filter. However, for the design of the WF controller, in the power tracking configuration, we can assume that $T_{g}\approx T_{g}^{ref}$ and $P_{out}\approx P_{ref}\approx P_{dem}$.
               
			\subsubsection{Linearized WT model} Defining the state $x^{WT}$, the input $u^{WT}$, the disturbance $d^{WT}$ and the output $y^{WT}$ as
               \begin{align*}
                 x^{WT}&=(\beta,~\omega_{r},~\omega_{g}^{f}) - (\beta_{ref0},~\omega_{r0},~\omega_{g0})\\
                 u^{WT}&= P_{dem} - P_{dem0}\\
                 d^{WT}&= v -v_{0}\\
                 y^{WT} &= (M_t,~M_s)-(M_{t0},~M_{s0}),
               \end{align*}
               the linearized dynamics is given by
               \begin{align}
                 \label{eq:SSState}\dot{x}^{WT}(t) &= A^{WT} x^{WT}(t) + B^{WT} u^{WT}(t)+ B_{d}^{WT} d^{WT}(t)\\
                 \label{eq:SSOutput}y^{WT}(t) &= C^{WT} x^{WT}(t) + D^{WT} u^{WT}(t)+ D_{d}^{WT} d^{WT}(t)
               \end{align}
               where matrices $A^{WT}$, $B^{WT}$, $B_{d}^{WT}$, $C^{WT}$, $D^{WT}$ and $D_{d}^{WT}$ are obtained from \eqref{eq:RotTorqLTI}-\eqref{eq:MsLTIpart} and \eqref{eq:Tglinear}.

          \subsection{Optimal one-step-ahead predictor of wind turbulence}
               The wind blowing on the wind farm generates an exogenous input $v$ that acts differently on each WT: for this reason, during the design phase of the controller, it is important to use as much as possible the knowledge of the wind field (see also \cite{Laks2011} and \cite{Biegel2011}). As common in the literature, we can rewrite the wind as $v = \bar{v} + \tilde{v}$ where $\bar{v}$ is an average speed, depending on weather conditions, which changes in the order of hours, and $\tilde{v}$ is a zero mean turbulence variation that varies on a faster time scale \cite[p. 17]{Burton2001}. This latter component of the wind is generally due to thermal conditions (e.g. variations in temperature) and the friction with the earth's surface. A wind profile is characterized by the average speed and the turbulence intensity $T_I$ defined as 
               \begin{equation*}
                 \label{eq:windTurbulence}
                 T_{I} = \frac{\sigma_{\tilde{v}}}{\bar{v}}
               \end{equation*}
               where $\sigma_{\tilde{v}}$ is the standard deviation. We can describe the turbulence spectrum as a function of frequency using the Kaimal spectrum (\cite[p. 23]{Burton2001} and \cite{Madjidian2011}), given by
               \begin{equation*}
                 \label{eq:kaimalSpec}
                 \Phi_{\tilde v} (\omega) = \sigma_{\tilde{v}}^{2} \frac{4 \frac{L_{v}}{\bar{v}}}{(1+ \omega \frac{3 L_{v}}{\pi \bar{v}} )^{\frac{5}{3}}}
               \end{equation*}
               where $L_{v}$ is a length scale $[m]$. In the literature, it is common to simplify the wind model by assuming the wind speed variations are distributed as a WGN. In addition, each WT in a WF is affected by the presence of neighboring WTs. This effect is neglected in this study and will be considered in future research.\\
               In order to derive a linearized model of a WF, a linear model of the wind is needed. We identify an ARMA model for the turbulence variation and then obtain an optimal predictor. The ARMA process is described by a linear combination of previous outputs $y^{ARMA}(t)$ and previous inputs $w_v(t)\sim WGN(0,\sigma^2)$ \cite{Ljung1999}. We can identify and validate an ARMA process for wind profiles described by each pair $(\bar{v},~T_I)$. Let the ARMA process described by the transfer function $G(z)=\frac{C(z)}{A(z)}$. Then, the optimal one-step-ahead predictor \cite{Ljung1999} can be derived as
               \begin{equation}
                 \label{eq:PzYz}
                 \hat{\tV}(z) = \frac{C(z)-A(z)}{C(z)} \tV(z)                 
               \end{equation}
               where $\tV(z)$ is the Z-transform of $\tv(t)$ and $\hat{\tV}(z)$ is the Z-transform of the predicted turbulence variation $\hat{\tv}(t|t-1)$. Minimal realization of \eqref{eq:PzYz} in the state-space yields to the model
               \begin{equation*}
                 \label{eq:windSS}
                 \begin{aligned}
                   x_v(t+1) &= A_v x_v(t) + B_v \tv(t)\\
                   \hat{\tv}(t|t-1) &= C_v x_v(t).
                 \end{aligned}
               \end{equation*}
               Moreover the prediction error is distributed as $w_v(t)$.
               
               \subsubsection{Importance of wind predictor} 
\begin{figure}[!htb]
                 \centering
                 \includegraphics[scale=0.35]{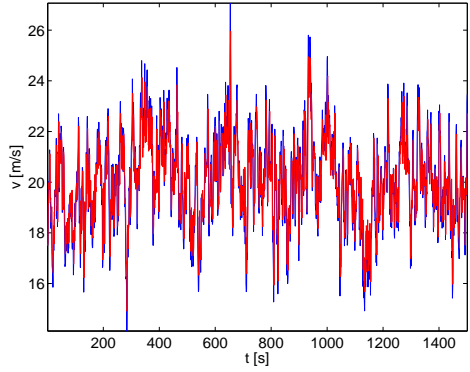}
                 \caption{Wind profiles for $v_0=20$ and  $T_I=0.1$: in blue the validation set and in red the predicted profile obtained using \eqref{eq:wind_profiles_20_01_test_1}.}
                 \label{fig:wind_profiles_20_01_test_1}
               \end{figure}

In the following, we show through examples the advantages of using an optimal one-step-ahead predictor for the wind turbulence. We consider two set of measurements of wind speed: the first set is used to identify the ARMA process and the second one is used to validate the optimal one-step-ahead predictor The sets have been produced using the SWF toolbox which allows one to generate wind profiles distributed according to the Kaimal spectrum. We identify ARMA processes by trying different combinations of previous measurements and Gaussian noise samples and we choose the optimal predictor that minimizes the Final Prediction Error (FPE). In the first example, we consider a wind speed described by $\bv=20$ and  $T_I=0.1$, hence $\sigma_{\tilde{v}}^2 = 4$. We obtain the following optimal predictor for the wind speed
               \begin{equation}
                 \label{eq:wind_profiles_20_01_test_1}
                 \begin{aligned}
                   x_v(t+1) &= \matr{ 0 & 0 & 0 \\ 0.25 & 0 & 0 \\ 0 & 0.5 & 0 } x_v(t) + \matr{1\\0\\0}(v(t)-20)\\
                   \hat{v}(t|t-1) &= \matr{ 0.9021 & -0.4406 & 0.5389 } x_v(t) + 20,
                 \end{aligned}
               \end{equation}
               where estimated variance of the prediction error is $0.9010$, i.e. it is distributed as $WGN(0,0.9010)$. The validation set and the predicted wind speed profiles are shown in Figure \ref{fig:wind_profiles_20_01_test_1}.
               
\begin{figure}[!htb]
                 \centering
                 \includegraphics[scale=0.35]{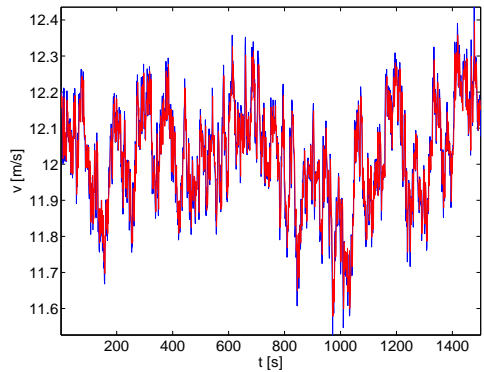}
                 \caption{Wind profiles for $v_0=12$ and  $T_I=0.01$: in blue the validation set and in red the predicted profile obtained using \eqref{eq:wind_profiles_12_001_test_1}.}
                 \label{fig:wind_profiles_12_001_test_1}
               \end{figure}

               In the second example, we consider a wind speed described by $\bv=12$ and  $T_I=0.01$, hence $\sigma_{\tilde{v}}^2 = 0.0144$. We obtain the following optimal predictor for the wind speed
               \begin{equation}
                 \label{eq:wind_profiles_12_001_test_1}
                 \begin{aligned}
                   x_v(t+1) &= \matr{ 0.3613 & 0.2621 \\ 0.5 & 0 } x_v(t) + \matr{1\\0}(v(t)-12)\\
                   \hat{v}(t|t-1) &= \matr{ 0.8885 & -0.8208 } x_v(t) + 12,
                 \end{aligned}
               \end{equation}
               where estimated variance of the prediction error is $0.0036$. The validation set and the predicted wind speed profiles are shown in Figure \ref{fig:wind_profiles_12_001_test_1}.
               In both examples, we note that variance of the original wind turbulence $\tv$ has been reduced by $75\%$. This is very useful in an MPC architecture, where we need to predict the behavior of each WT and hence to predict also the wind turbulence over a time horizon.

          \subsection{WF model}
               In this section, we derive a model of the WF. First, we derive a local model of a WT. We note that the dynamics in \eqref{eq:SSState} and \eqref{eq:SSOutput} depend on the measurements of the wind turbulence. However, using MPC, we need to predict the wind turbulence. To this purpose, we will use the optimal predictor obtained in the previous section by setting $\bv = v_0$ where $v_0$ is the wind speed at the operating point (see Section \ref{subsub:aereo}).\\
               Discretizing dynamics in \eqref{eq:SSState} and \eqref{eq:SSOutput} with sampling time $T_s=1$ sec\footnote{Note that the choice of the sampling time depends on the working frequency of the WF SCADA system, that is usually 1Hz.}, augmenting the state of the WT using the states of the optimal predictor and using the fact that 
               $$d^{WT}(t)=\hat{\tv}(t|t-1)+w_v(t)=\tv(t),$$ 
               we obtain the following discrete-time LTI model
               \begin{equation}
                 \label{eq:WTwindSS1}
                 \begin{aligned}
                   x^{WT}_a(t+1) &= A_a x^{WT}_a(t) + B_a u^{WT}(t) + B_{da} w_v(t)\\
                   y^{WT}(t) &= C_a x^{WT}_a(t) + D_a u^{WT}(t) + D_{da} w_v(t)
                 \end{aligned}
               \end{equation}
               where
               \begin{equation*}
                 x^{WT}_a = (x^{WT},~x_v)
               \end{equation*}               
               \begin{align*}
                 &A_a = \begin{bmatrix}
                   \bar{A}^{WT} & \bar{B}_{d}^{WT}C_{v}\\
                   0 & A_v+B_vC_v
                 \end{bmatrix},~
                 B_a = \begin{bmatrix}
                   \bar{B}^{WT} \\
                   0 \end{bmatrix},~
                 B_{da} = \begin{bmatrix}
                   \bar{B}_{d}^{WT} \\
                   B_v\end{bmatrix}\\
                 &C_a = \begin{bmatrix}
                   \bar{C}^{WT} & \bar{D}_{d}^{WT}C_v
                 \end{bmatrix},~
                 D_a =\bar{D}^{WT},~
                 D_{da} =\bar{D}_d^{WT}
               \end{align*}
               and $\bar{A}^{WT}$, $\bar{B}^{WT}$, $\bar{B}_{d}^{WT}$, $\bar{C}^{WT}$, $\bar{D}^{WT}$ and $\bar{D}_d^{WT}$ are discrete-time counterparts (obtained through exact discretization) of the corresponding matrices in \eqref{eq:SSState} and \eqref{eq:SSOutput}. 

               In order, to derive a WF model consisting of $N$ turbines, we need to group $N$ WT models described by \eqref{eq:WTwindSS1}. Therefore the WF model\footnote{With abuse of notation, the state, input, output and disturbance variables of the $i$-th WT, as well as matrices, are indicated with subscript $i$.} is given by
               \begin{equation}
                 \label{eq:WFSS1}
                 \begin{aligned}
                   x(t+1) &= A x (t) + B u(t) + B_d w(t)\\
                   y(t) &= C x(t) + D u(t)+D_d w(t)
                 \end{aligned}
               \end{equation}
               where
               \begin{align*}
                 x &= ( x^{WT}_{a,1},\ldots,x^{WT}_{a,N} ),~ &u = ( u^{WT}_{1},\ldots,u^{WT}_{N})\\
                 y &= ( y^{WT}_{1},\ldots,y^{WT}_{N}),~&w =( w_{v,1},\ldots,w_{v,N})\\
                 w &\sim WGN(0,\Sigma_{w}),~&\Sigma_{w} = \diag(\sigma_{1}^{2},\ldots,\sigma_{N}^{2})\\
                 A &= \diag(A_{a,1},\ldots,A_{a,N}),~&B = \diag(B_{a,1},\ldots,B_{a,N})\\
                 B_{d} &= \diag(B_{da,1},\ldots,B_{da,N}),~&C = \diag(C_{a,1},\ldots,C_{a,N})\\
                 D &= \diag(D_{a,1},\ldots,D_{a,N}),~&D_d= \diag(D_{da,1},\ldots,D_{da,N}).
               \end{align*}
               In the following section, we will use model \eqref{eq:WFSS1} to predict the behavior of the WF. Moreover we will detail how we can take into account the wind measurements at each time-instant.

     \section{MPC regulators for WFs}
          \label{sec:mpcpre}
          In this section we present different MPC regulators for achieving optimal power dispatching.
 We first introduce performance measures for assessing the quality of a dispatching algorithm. Then, we present the basic MPC formulation with chance constraints and, finally, we derive SMPC and DMPC regulators.        
          \subsection{Performance measures}
               \label{sec:costfunction}
               In order to evaluate performance of different regulators, quantitative criteria are needed. In this paper, we use the index $\tJ$ proposed in \cite{Couchman2008}
               \begin{equation}
                 \label{eq:costFunctionVestas}
                 \tJ = J_P + J_{M_s} + J_{M_t}
               \end{equation}
               where
               \begin{itemize}
               \item $J_P$ is a measure of the power production and it is defined as
                 $$
                 J_P = \sqrt{\frac{1}{T}\int_0^T\frac{1}{NP_{rated}}\sum_{i=1}^{N}\left(P_i(t)-P_{ref,i}(t)\right)^2 dt}
                 $$
                 with $P_{rated}$ is the wind turbine rated power;
               \item $J_{M_s}$ is a measure of the total main shaft fatigue and is defined as
                 \begin{equation}
                 	\label{eq:JMS}
                 	J_{M_s} = \sum_{i=1}^N 0.2~std\left[\frac{M_{s,i}(0:T)}{2\cdot 10^6}\right]
                 \end{equation}
               \item $J_{M_t}$ is a measure of the fore-aft oscillation on the tower and is defined as
                 \begin{equation}
                 	\label{eq:JMT}
                 	J_{M_t} = \sum_{i=1}^N 0.05~std\left[\frac{M_{t,i}(0:T)}{23\cdot 10^6}\right].
                 \end{equation}   
               \end{itemize}
               Note that in $J_{M_s}$ and $J_{M_t}$ we use the standard deviation instead of using the rain-flow algorithm and Palmgren-Miner sum as in \cite{Fuchs1980}. This is due to the fact that the working frequency of a WF SCADA system is not high enough to represent the damage fatigue obtained with a rain-flow count. This problem has been widely investigated in literature and we defer the interested readers to \cite{Stephens2001,Couchman2008,Knudsen2011a}.

          \subsection{MPC formulation}
               \label{sec:mpcformulation}
               In the following, we use $x_t$, $u_t$, $y_t$, $w_t$ and $d_t$ instead of $x(t)$, $u(t)$, $y(t)$, $w(t)$ and $d(t)$, respectively. At each time instant $t$, we solve the following MPC optimization problem over the prediction horizon $N_h$
               \begin{subequations}
                 \label{prbl:Original}
                 \begin{align}
                  \min_{\substack{u_k,~\forall k= t:t+N_h}}\qquad \sum_{k=t}^{t+N_h} \Eset\left[\norme{y_{k}}{Q}^2 + \norme{u_{k}}{R}^2\right]\label{eq:costMPC}\\
                   x_{k+1} = A x_{k} + B u_{k} + B_{d} w_{k},~\forall k= t:t+N_h, \label{eq:xconstr}\\
                   y_{k} = C x_{k} + D u_{k}+ D_{d} w_{k},~\forall k= t:t+N_h, \label{eq:yconstr}\\
                   \One_N^{T} u_k = 0,~\forall k= t:t+N_h, \label{eq:uconstr}\\
                   \PP(c_{s}^T u_k \geq u_{s}^{max}) \leq \tp,\;\;\; \forall s = 1:S, ~\forall k= t:t+N_h. \label{eq:uconstrdis}
                 \end{align}
               \end{subequations}
For short, in \eqref{prbl:Original} we used the index $k$ (instead of the double index $k,t$) for referring to variables within the prediction horizon $t:t+N_h$.
               Note that cost function \eqref{eq:costMPC} and constraint \eqref{eq:uconstr} correspond to minimize the total load fatigue while guaranteeing that the total power generated by the WF fulfills the power demand required by the network operator. Moreover inequalities \eqref{eq:uconstrdis} represent linear probabilistic input constraints, where $c_{s}\in\Rset^N$, $u_s^{max}>0$ (so that the input constraint is inactive for $u_k=0$), $S$ is the number of linear input constraints and $\tp$ is a maximal probability of constraint violation. In view of \eqref{eq:JMS} and \eqref{eq:JMT}, we set 
               \begin{equation*}
                 Q=\diag\left(\matr{q_{M_t} & 0 \\ 0 & q_{M_s}},\ldots,\matr{q_{M_t} & 0 \\ 0 & q_{M_s}}\right),~\mbox{with }q_{M_t}=\frac{0.05}{(23\cdot 10^6)^2N_{h}}\mbox{ and }q_{M_s}=\frac{0.2}{(2\cdot 10^6)^2N_{h}}.
               \end{equation*}
               Furthermore, we assume $R=\diag(r_1,\ldots,r_N)$ and hence the only tunable parameters are $r_i>0,~\forall i=1:N$.
               
               In order to remove constraint \eqref{eq:uconstr} following \cite[p. 537]{Boyd2004} (see also \cite{Biegel2011} and \cite{Madjidian2011}), we look for a matrix $T\in \Rset^{N\times N-1}$ that parameterizes the linear feasible set
               \begin{equation*}
                 \{u_{k} \in \Rset^N: \One_N^{T} u_{k} = 0 \} = \{ T \hu_k\ : \hu_k \in \Rset^{N-1}\}.
               \end{equation*}
This can be achieved using the following transformation in the input space
               \begin{equation}
                 \label{eq:Tuhat}
                 u_{k} = T\hat{u}_{k}                 
               \end{equation}
               where
               \begin{equation*}
                 T = 
                 \begin{bmatrix}
                   1 & 0 & \dots & 0\\
                   -1 & 1 & \ddots & \vdots \\
                   \vdots & \ddots & \ddots & 0 \\
                   0 & \dots & -1 & 1
                 \end{bmatrix}.
               \end{equation*}

               Differently from \cite{Madjidian2011}, \cite{Biegel2011} and \cite{Biegel2013a}, next we show how to take into account wind measurements at time instant $t$. Since the value $d^{WT}(t)$ in \eqref{eq:SSState} and \eqref{eq:SSOutput} is known for each WT, constraints \eqref{eq:xconstr} and \eqref{eq:yconstr} for $k=t$ can be rewritten as
               \begin{equation}
                 \label{eq:WFSS2}
                 \begin{aligned}
                   x_{t+1} &= A_0 x_t + B u_t + B_d d_t\\
                   y(t) &= C_0 x_t + D u_t+D_d d_t
                 \end{aligned}
               \end{equation}
               where
               \begin{align*}
                 d &= ( d^{WT}_{1},\ldots,d^{WT}_{N})\\
                 A_0 &= \diag(A_{0,1},\ldots,A_{0,N}),~&C = \diag(C_{0,1},\ldots,C_{0,N})\\
                 &A_{0,i} = \begin{bmatrix}
                   \bar{A}^{WT} & 0 \\
                   0 & A_v
                 \end{bmatrix},~&C_{0,i} = \begin{bmatrix}
                   \bar{C}^{WT} & 0
                 \end{bmatrix}.
               \end{align*}               
               Therefore, in \eqref{eq:xconstr} the state $x_{t+1}$ depends in a deterministic way on $x_t$, $u_t$, since $w_t$ is fixed.

               Summarizing, using \eqref{eq:yconstr}, \eqref{eq:Tuhat} and \eqref{eq:WFSS2}, we can rewrite the MPC optimization problem as
               \begin{subequations}
                 \label{prbl:mpcbase}
                 \begin{align}
                   \min_{\substack{\hu_k,~\forall k= t:t+N_h}}\qquad \sum_{k=t}^{t+N_h} \Eset\left[\norme{y_{k}}{Q}^2 + \norme{\hu_{k}}{\hR}^2\right]\label{eq:costMPC_base}\\
                   x_{t+1} = A_0 x_{t} + \hB \hu_{t} + B_{d} d_{t}, \label{eq:xconstr_base_0}\\
                   y_{t} = C_0 x_{t} + \hD \hu_{t}+ D_{d} d_{t},\label{eq:yconstr_base_0}\\
                   x_{k+1} = A x_{k} + \hB \hu_{k} + B_{d} w_{k},~\forall k= t+1:t+N_h, \label{eq:xconstr_base_t}\\
                   y_{k} = C x_{k} + \hD \hu_{k}+ D_{d} w_{k},~\forall k= t+1:t+N_h, \label{eq:yconstr_base_t}\\
                   \PP(\hc_{s}^T \hu_k \geq u_{s}^{max}) \leq \tp,\;\;\; \forall s = 1:S, ~\forall k= t:t+N_h, \label{eq:uconstrdis_base}
                 \end{align}
               \end{subequations}
               where
               $$
               \hB = BT,\qquad \hD = DT,\qquad \hR = T^TRT,\qquad \hc_{s}^T = c_s^TT.
               $$
			   Note that constraint \eqref{eq:xconstr} (resp. \eqref{eq:yconstr}) has been split into constraints \eqref{eq:xconstr_base_0} and \eqref{eq:xconstr_base_t} (resp. \eqref{eq:yconstr_base_0} and \eqref{eq:yconstr_base_t}).

          \subsection{SMPC regulator}
               In this section we design an SMPC regulator. Our aim is to rewrite the stochastic problem \eqref{prbl:mpcbase} as a deterministic optimization problem solvable through Semi-Definite Programming (SDP) \cite{Boyd2004}. To this purpose we adopt the approach to SMPC proposed in \cite{Magni2009a} and \cite{Farina2013b}.

The optimization problem that must be solved online at each time instant $t$ is
               \begin{equation}
                 \label{prbl:mpcbase_stochastic}
                 \begin{aligned}
                   \min\qquad tr\left(M_0P_t \right) + \sum_{k=t+1}^{t+N_h}  tr\left(M P_{k} \right)
 \end{aligned}
               \end{equation}
with respect to the unknowns $\hat U_k$, $G_k$ $P_k$, $\bar\hu_{k}$, for $k= t:t+N_h$, $\bar x_{k}$, for $k= t+1:t+N_h$, $\theta_{ks}$ for $k= t:t+N_h$ and $s = 1:S$, $X_k$ for  $k= t+3:t+N_h$ and  subject to the LMI constraints
{\small
\begin{equation}
\begin{aligned}
\bx_{t+1} &= A_0 x_t + \hB \bar\hu_{t} + B_{d} d_{t} \\
  \bx_{k+1} &= A \bx_{k} + \hB \bar\hu_{k},~\forall k= t+1:t+N_h   \label{eq:nominalDyn}
\end{aligned}
 \end{equation}
\begin{equation}
                   X_{t} = X_{t+1} = 0 \label{eq:x0x1}
                 \end{equation}
\begin{equation}
                   X_{t+2}= B_{d} \Sigma_{w} B_{d}^{T}.\label{eq:x2}
                 \end{equation}
\begin{equation}
                 \label{eq:LMIvarianceDyn}
                 \matr{
                   X_{k+1} & AX_{k}+\hat{B}G_{k} & B_{d}\Sigma_{w}\\
                   (\ast) & X_{k} & 0 \\
                   (\ast) & (\ast) & \Sigma_{w}
                 } \geq 0,\qquad X_{k}\geq 0, \quad k=t+3:t+N_h
               \end{equation}
\begin{align}
                 \label{eq:Pk}
                 \begin{bmatrix}
                   P_{k} & \matr{ X_{k} & 0 \\ G_{k} & 0 \\ 0 & \Iset } & \matr{\bx_k\\\bar\hu_k\\0} \\
                   (\ast) & \matr{ X_{k} & 0 \\ 0 & \Sigma_w^{-1} } & 0 \\
                   (\ast) & (\ast) & 1
                 \end{bmatrix},&\geq 0 &\forall k = t+2, \ldots, N_h 
               \end{align}
              
               \begin{align}
                 \label{eq:Pj}
                 \begin{bmatrix}
                   P_{j} & \begin{bmatrix} \bar{x}_{j} \\ \bar{\hat{u}}_{j} \\0 \end{bmatrix} \\
                   (\ast)& 1
                 \end{bmatrix}&\geq 0
               \end{align}
\begin{align}                 
                 &\label{eq:LMIinputcons1} \begin{bmatrix} \hat{U}_{k} & G_{k}\\
                   G_{k}^{T} & X_{k} \end{bmatrix} \geq 0,\quad k = t:t+N_h\\
                 &\label{eq:LMIinputcons2} \hat{c}_{s}^T \bar{\hat{u}}_{k} \leq \frac{3}{4} u_{s}^{max} - \frac{\theta_{ks}}{u_{s}^{max}},\quad k = t:t+N_h \text{ and } s = 1:S \\
& \label{eq:LMIinputcons2bis}\theta_{ks}>0 ,\quad k = t:t+N_h \text{ and } s = 1:S \\
                 &\label{eq:LMIinputcons3} \hat{c}_{s}^{T} \hat{U}_{k} \hat{c}_{s} \leq \theta_{ks} \frac{1}{2} \left( \frac{1}{erf^{-1}(1 - 2\tp)} \right)^{2},\quad k = t:t+N_h \text{ and } s = 1:S
               \end{align}
}
where $(\ast)$ denotes the matrix transpose of the corresponding block in the upper triangular part, $erf(\cdot)$ is the Gauss error function and
{\small
where 
               $$
               M_0 = \matr{ C_0^TQC_0 & C_0^TQ\hD & C_0^TQD_d \\  (\ast) & T^TRT+\hD^TQD & \hD^TQD_d \\  (\ast) &  (\ast) & D_d^TQD_d},
               \quad
               M = \matr{ C^TQC & C^TQ\hD & C^TQD_d \\  (\ast) & T^TRT+\hD^TQD & \hD^TQD_d \\  (\ast) &  (\ast) & D_d^TQD_d}.
               $$
}  

			   The control law $\hat{u}_t$ is then obtained as 
			   \begin{equation}
                 \label{eq:controlLawt}
                 \hat{u}_{t} = \bar{\hat{u}}_{t}.
               \end{equation}

				A detailed derivation of problem \eqref{prbl:mpcbase_stochastic} from problem \eqref{eq:costMPC_base} is described in the Appendix.
Here, we just highlight that, up to the linearization of a square-root function which is needed for getting the affine constraint \eqref{eq:LMIinputcons2}, feasibility of \eqref{eq:nominalDyn}-\eqref{eq:LMIinputcons3} implies that chance constraints \eqref{eq:uconstrdis_base} are fulfilled. Moreover, the cost in \eqref{prbl:mpcbase_stochastic} provides an upper bound to the cost in \eqref{eq:costMPC_base}. As shown in \cite{Magni2009a}, tightening of the constraints \eqref{eq:uconstrdis_base} and relaxation of the cost in \eqref{eq:costMPC_base} are needed for recasting the original nonlinear optimization problem into an SDP problem.

          \subsection{DMPC regulators}
               In order to design a DMPC regulator, we do not consider stochasticity in the optimization problem \eqref{prbl:mpcbase}, hence $w_k=0$, $\forall k=t+1:t+N_h$, . Therefore, the MPC problem can be rewritten as
               \begin{subequations}
                 \label{prbl:mpcbase_deterministic}
                 \begin{align}
                   \min_{\bar\hu_k,~\forall k= t:t+N_h}\qquad \sum_{k=t}^{t+N_h} \norme{\by_{k}}{Q} + \norme{{\bar\hu}_{k}}{\hR} + \norme{\epsilon_{k,1:S}}{\rho} \label{eq:costMPC_base_deterministic}\\
                   \bx_{t+1} = A_0 x_{t} + \hB \bar\hu_{t} + B_{d} d_{t}, \label{eq:xconstr_base_0_deterministic}\\
                   \by_{t} = C_0 x_{t} + \hD \bar\hu_{t}+ D_{d} d_{t},\label{eq:yconstr_base_0_deterministic}\\
                   \bx_{k+1} = A \bx_{k} + \hB \bar\hu_{k},~\forall k= t+1:t+N_h, \label{eq:xconstr_base_t_deterministic}\\
                   \by_{k} = C \bx_{k} + \hD \bar\hu_{k},~\forall k= t+1:t+N_h, \label{eq:yconstr_base_t_deterministic}\\
                   \hc_s^T\bar\hu_k\leq u_s^{max} + \epsilon_{k,s} ,~\forall k= t:t+N_h,~\forall s=1:S, \label{eq:uconstrdis_base_deterministic}\\
                   \epsilon_{k,s}\geq 0 ,~\forall k= t:t+N_h,~\forall s=1:S \label{eq:epsilonconstrdis_base_deterministic}
                 \end{align}
               \end{subequations}
               where the bar on a variable denotes the mean value. Moreover, we replace probabilistic constraints \eqref{eq:uconstrdis_base} with linear constraints \eqref{eq:uconstrdis_base_deterministic} where we introduced the slack variables $\epsilon_{k,s}$. Slack variables are also weighted in the cost function \eqref{eq:costMPC_base_deterministic}, where we assume $\rho>0$. Using a deterministic MPC regulator, we have to solve a QP problem at each time instant: from a computational point of view, even if the order of the LTI system \eqref{eq:WFSS1} increases, the optimization problem can be solved online with high sampling rate \cite{Boyd2004}. Moreover, in absence of constraint on the inputs, constraint \eqref{eq:uconstrdis_base_deterministic} do not appear in the optimization problem \eqref{prbl:mpcbase_deterministic}. Hence, the optimal value of slack variables is $\epsilon_{k,s}=0 ,~\forall k= t:t+N_h,~\forall s=1:S$, and we can solve \eqref{prbl:mpcbase_deterministic} explicitly, obtaining
               \begin{equation}
                 \label{eq:umpcdeterministic}
                 \matr{ \bar\hu_t \\ \bar\hu_{t+1} \\ \vdots \\ \bar\hu_{t+N_h}  } = -(\BB^T\QQ\BB+\RR)^{-1}\left(\BB^T\QQ\AAA x_t + \BB^T\QQ\BB_d d_t\right)
               \end{equation}
               where
               $$
               \BB = \matr{ \hD & 0 & 0 & 0 & \cdots & 0  \\ C\hB & \hD & 0 & 0 & \cdots & 0 \\ CA\hB & C\hB & \hD & 0 & \cdots & 0 \\ \vdots & \vdots & \vdots & \ddots & \ddots & \vdots \\ CA^{N_h-1}\hB & CA^{N_h-2}\hB & \cdots & \cdots & C\hB & \hD },~
               \AAA = \matr{ C_0 \\ CA_0 \\ CAA_0 \\ \vdots \\ CA^{N_h-1}A_0 },~\BB_d = \matr{ D_d \\ CB_d \\ CAB_d \\ \vdots \\ CA^{N_h-1}B_d }
               $$
               $$
               \QQ = \diag(Q,\ldots,Q),\qquad \RR = \diag(R,\ldots,R).
               $$    
               In the sequel, we will refer to this approach as Explicit DMPC (EDMPC). We note that in \eqref{eq:umpcdeterministic} the control inputs over the prediction horizon depend both on the measured state $x_t$ and the wind turbulence measurements $d_t$. Furthermore, \eqref{eq:umpcdeterministic} can be easily implemented in a SCADA system without requiring optimization tools. On the other hand, since the input constraints are not involved in the MPC problem, the matrix $R$ must be chosen properly, as we will show in the example section.

          \subsection{On-line control actions}
               Summarizing, at each time instant $t$, the power demand set-points for the WTs are computed as
               \begin{equation*}
                 P_{dem}(t) = T\hu^{MPC}(t) + P_{dem0}
               \end{equation*}
               where $P_{dem}(t)=(P_{dem,1}(t),~\ldots,~P_{dem,N}(t))$, $P_{dem0}=(P_{dem0,1},~\ldots,~P_{dem0,N})$ and, using the receding horizon principle, $\hu^{MPC}(t)$ is the optimal value $\bar\hu_t$ obtained
               \begin{itemize}
               \item solving the SDP optimization \eqref{prbl:mpcbase_stochastic} for the SMPC regulator
               \item solving the QP optimization \eqref{prbl:mpcbase_deterministic} for the DMPC regulator
               \item computing the control inputs \eqref{eq:umpcdeterministic} for the EDMPC regulator.
               \end{itemize}

		\subsection{SWF controller}
			In the simulation examples, we will compare performance of the proposed controller to the controller provided with the SWF toolbox (in the following SWFctrl). SWFctrl dispatches $P_{dem}^{WF}$ between the turbines proportionally to the available power at each turbine. In particular, the controller is based on the following equation
          \begin{align}
            P_{a} = \sum P_{a,i},\qquad P_{a,i} = \frac{\pi}{2} \rho R^{2} v_{meas,i}^{3} C_{p_{i}}^{max}\label{eq:pdistrSimWF}            
          \end{align} 
          where $P_{a,i}$ is the available power, $v_{meas,i}$ is the measured wind speed and $C_{p_{i}}^{max}$ is the maximum power coefficient for the $i$-th WT. Therefore, the WF power demand is distributed as
          \begin{equation*}
            P_{dem,{i}} = P_{dem}^{WF} \frac{P_{a,i}}{P_{a}}.
          \end{equation*}
          Note that \eqref{eq:pdistrSimWF} is the maximal value of $P_{a}$ that can be obtained in \eqref{eq:windTurbPow}.

     \section{Simulation examples}
          \label{sec:simresults}
          In order to assess the performance of the proposed control schemes, we use the SWF toolbox \cite{Grunnet2010}. The SWF toolbox allows to simulate a WF scenario using the Taylor's frozen turbulence hypothesis. This hypothesis, illustrated more thoroughly in \cite{Davidson2004}, concerns the interactions among WT due to the wind. In all simulations we will not make use of simplifications introduced by Taylor's hypothesis, therefore we do not introduce simplifications in generating an ambient wind field and we not reduce the complexity of wake effect models \cite{Grunnet2010}. Moreover, a WF modeled using the SWF toolbox presents further nonlinearities that were not considered in the design of our controller, such as elastic forces for the tower bending moment and saturation for pitch actuator and WT local controller. In the following, we show the simulations performed using MatLab/Simulink. In order to solve online the SDP and QP problems, we have used YALMIP \cite{Lofberg2004} and MOSEK \cite{MOSEKApS2013}.

          \subsection{WF composed of $10$ WTs as in \cite{Grunnet2010}}
          	   \label{sec:WT10}
\begin{figure}[!htb]%
                 \centering
                 \includegraphics[scale=0.42]{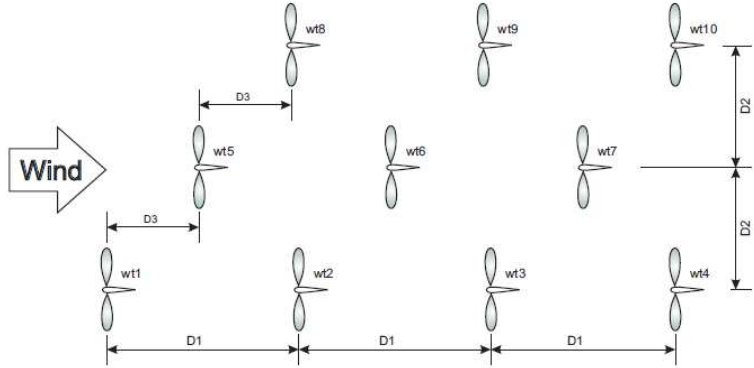}%
                 \caption{WF layout for example as in \cite{Grunnet2010}. $D1 = 600\;[m]$, $D2 = 500\;[m]$ and $D3 = 300\;[m]$. (Figure from \cite{Grunnet2010}).}%
                 \label{fig:WF10}%
               \end{figure}

               In the first example, we test our control architectures for a WF composed of $10$ WTs arranged as shown in Figure \ref{fig:WF10} and proposed in \cite{Grunnet2010}.
               For this example, we have set a WF power demand $P_{dem}^{WF} = 30\;[MW]$, equally distributed by the scheduler on the $10$ WTs, hence $P_{dem0,{i}} = 3\;[MW]$, $i=1:10$. Moreover, in the MPC cost function, we set $r_i=0.06$ for SMPC and DMPC and $r_i=0.1$ for EDMPC and we require that $\abs{u_i(t)}\leq 0.1~[MW]$. The parameters $r_i$ and the constraints on $u_i(t)$ are set in order to guarantee good performance around the given set-point $P_{dem0,i}$. Indeed, far from the set-point, predictions using a linearized model could be inaccurate. Moreover, the power demand set-points can be changed accordingly with the limitations given by the SCADA system. The prediction horizon is $N_h=2$. The wind speed at the operating point is $v_{0} = 12\;\left[ \frac{m}{s} \right]$ and its turbulence is $T_I=0.1$. The state-space model of the wind optimal predictor, used for all WTs, is
               \begin{equation}
                 \begin{aligned}
                   x_{i}^{w}(t+1) &= \begin{bmatrix} 0.7039 & 0.1116 \\ 0.5 & 0 \end{bmatrix} x_{i}^{w}(t) + \begin{bmatrix} 2 \\ 0 \end{bmatrix} (v_{meas_{i}}(t) - v_{0})\\
                   d_{i}^{w}(t) &= \begin{bmatrix} 0.4189 & -0.6178 \end{bmatrix} x_{i}^{w}(t),
                   \label{eq:windSSEx2}
                 \end{aligned}
               \end{equation}
               with variance of the prediction error equal to $0.3512$.\\
               In Table \ref{tab:exampleWT10}, we summarize performance using different controllers. We note that using the SWFctrl we achieve better performance in terms of tracking of the required power, however SWFctrl induces more mechanical stress, in particular for the main shaft. MPC schemes improve performance in terms of mechanical stress: indeed, compared with the open-loop controller, using MPC controllers we can improve performance at least of $19.69\%$ for $M_s$ and $3.29\%$ fo $M_t$. We also highlight that performance of DMPC and SMPC are better than using EDMPC: this is due to the fact that the weights $r_i$ are higher for EDMPC in order to guarantee that the power demand for each WT does not change more than $0.1~[MW]$. Performance of DMPC and SMPC are comparable: however, solving a QP has computational burden lower than solving an SDP.
               \begin{table}[!htb]
                 \centering
                   \begin{tabular}{|c||c||c||c||c|}
                     \hline
                     & $\tJ$ & $J_P$ & $J_{M_s}$ &  $J_{M_t}$ \\
                     \hline
                     Scheduler only    &      0.2551    &   0.0027    &    0.0611    &     0.1913      \\
                     \hline
                     SWFctrl & -140.77\% &  \textbf{10.64\%}  &   -574.71\%   &  -4.43\% \\
                     \hline
                     EDMPC & 7.26\% &  7.44\%  &   19.69\%   & 3.29\% \\
                     \hline
                     DMPC & \textbf{8.96\%} &  5.81\%  &   \textbf{22.30\%}   &  \textbf{4.75\%} \\
                     \hline
                     SMPC & 8.46\% &  4.78\%  &  21.19\%   &  4.45\% \\
                     \hline
                   \end{tabular}
                 \caption{Controllers performance for a WF composed of $10$ WTs as in \cite{Grunnet2010}. Table entries have been obtained by averaging values obtained in $5$ simulations of $15$ minutes each. Top row: performances using open-loop scheduling. Other rows: percentage increment/decrement with respect to the values in the first row. Best performances are in bold. }
                 \label{tab:exampleWT10}
               \end{table}

\begin{figure}[!htb]
                 \centering
                 \begin{subfigure}[!htb]{0.45\textwidth}
                   \centering
                   \includegraphics[scale=0.3]{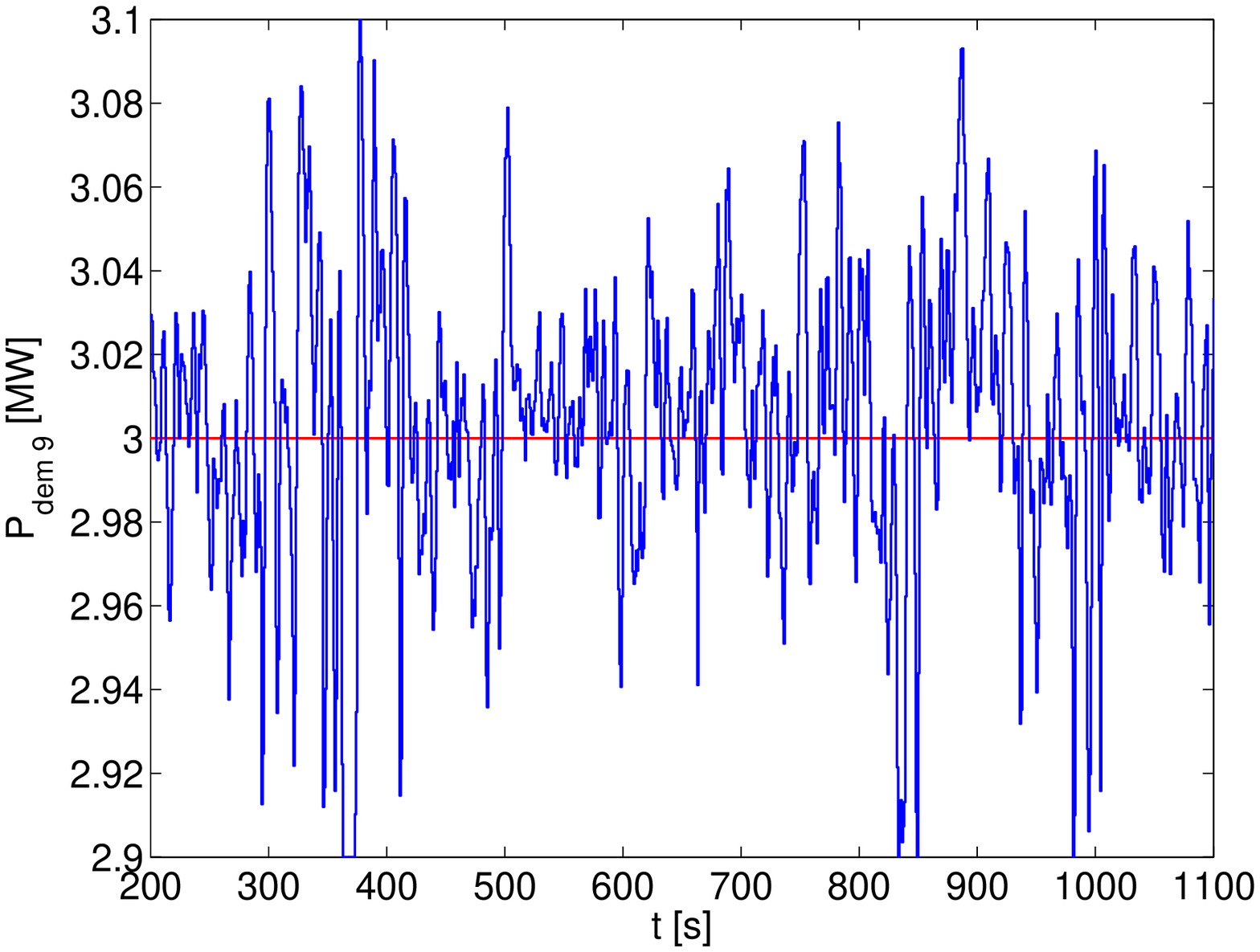}
                   \caption{Power demand for WT $9$.}
                   \label{fig:WT9powerdemand}
                 \end{subfigure}
                 \begin{subfigure}[!htb]{0.45\textwidth}
                   \centering
                   \includegraphics[scale=0.3]{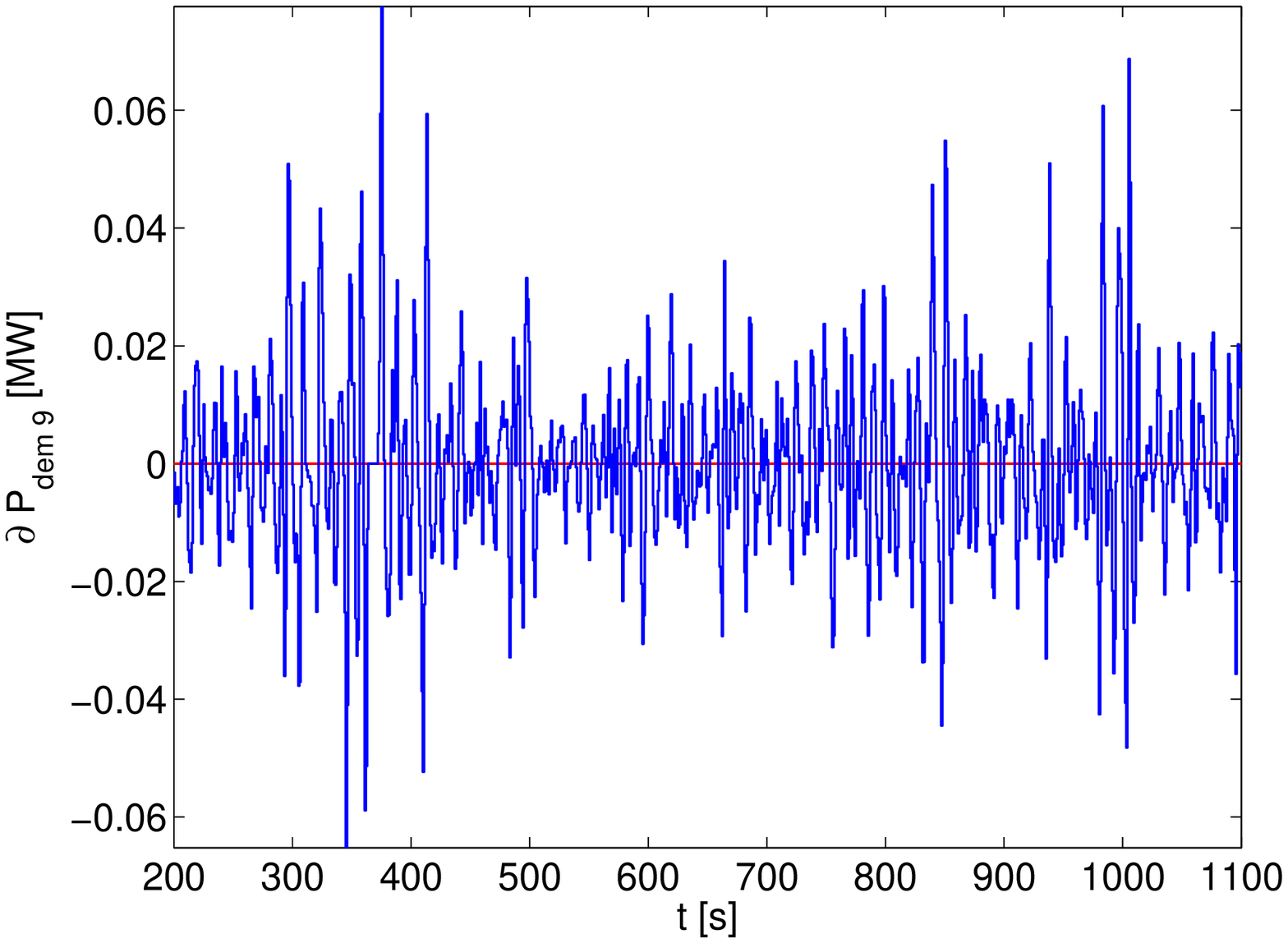}
                   \caption{Gradient of power demand for WT $9$.}
                   \label{fig:WT9gradientpowerdemand}
                 \end{subfigure}\\
                 \begin{subfigure}[!htb]{0.45\textwidth}
                   \centering
                   \includegraphics[scale=0.3]{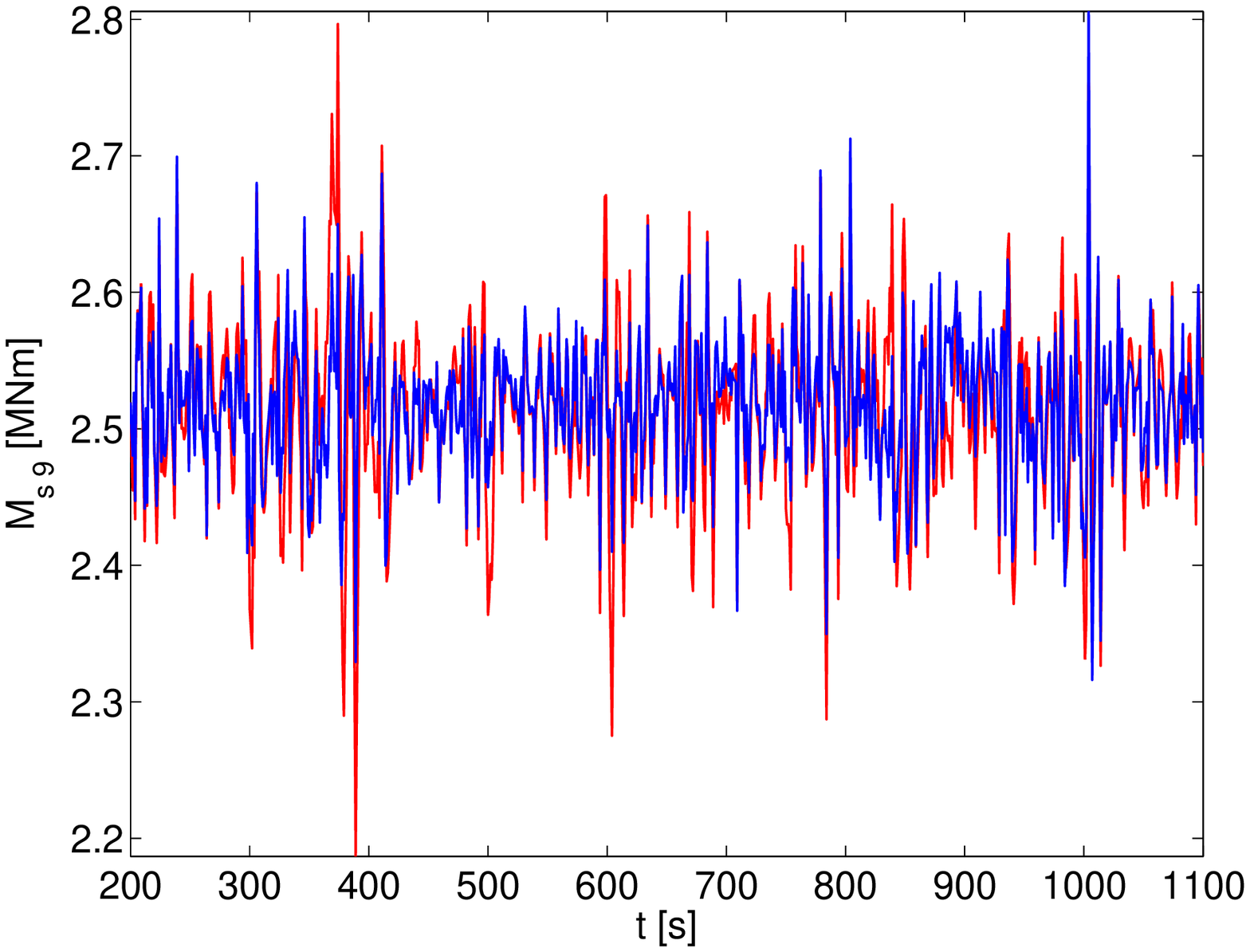}
                   \caption{Main shaft for WT $9$.}
                   \label{fig:WT9mainshaft}
                 \end{subfigure}
                 \begin{subfigure}[!htb]{0.45\textwidth}
                   \centering
                   \includegraphics[scale=0.3]{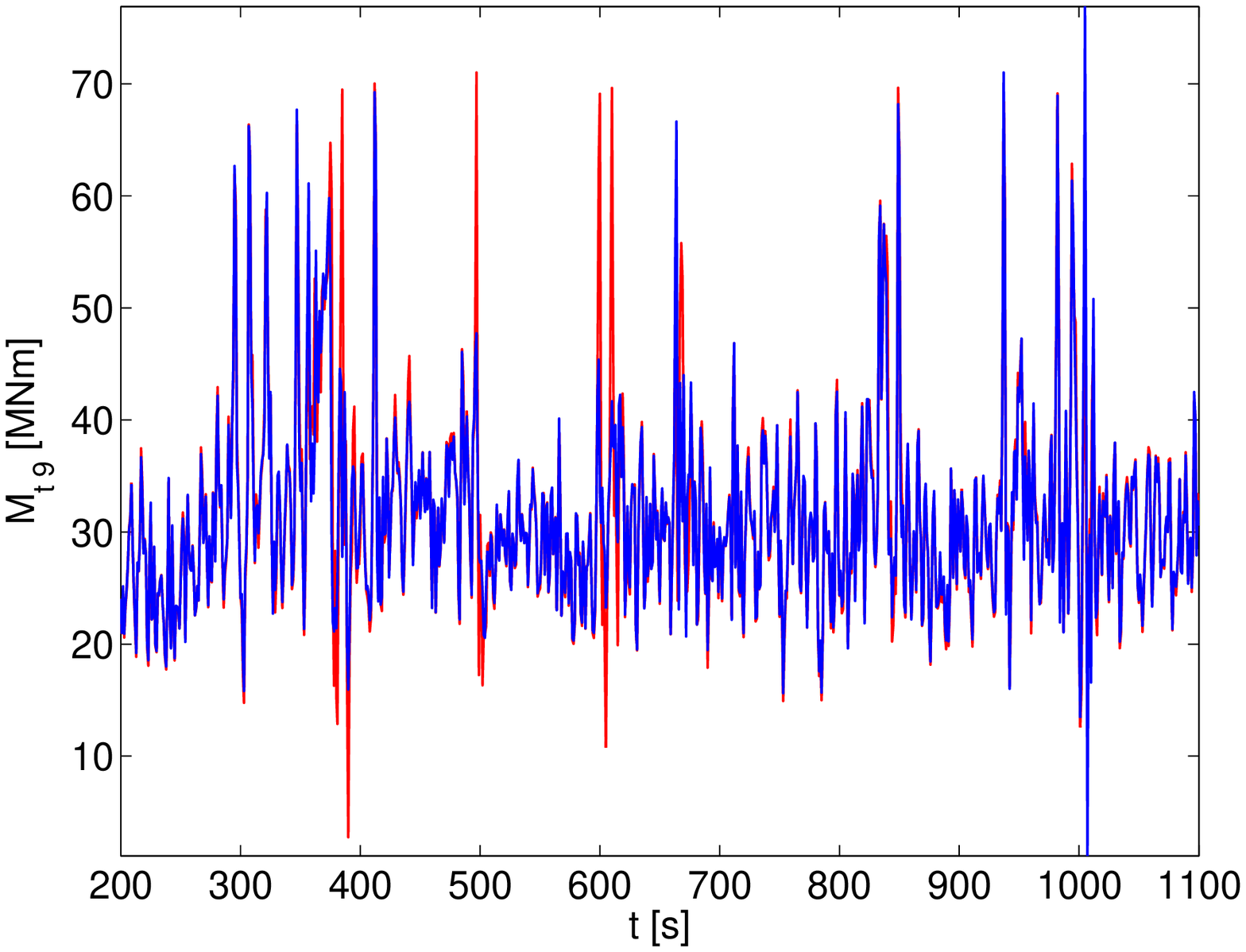}
                   \caption{Bending moment for WT $9$.}
                   \label{fig:WT9momentTower}
                 \end{subfigure}\\
                 \begin{subfigure}[!htb]{0.45\textwidth}
                   \centering
                   \includegraphics[scale=0.3]{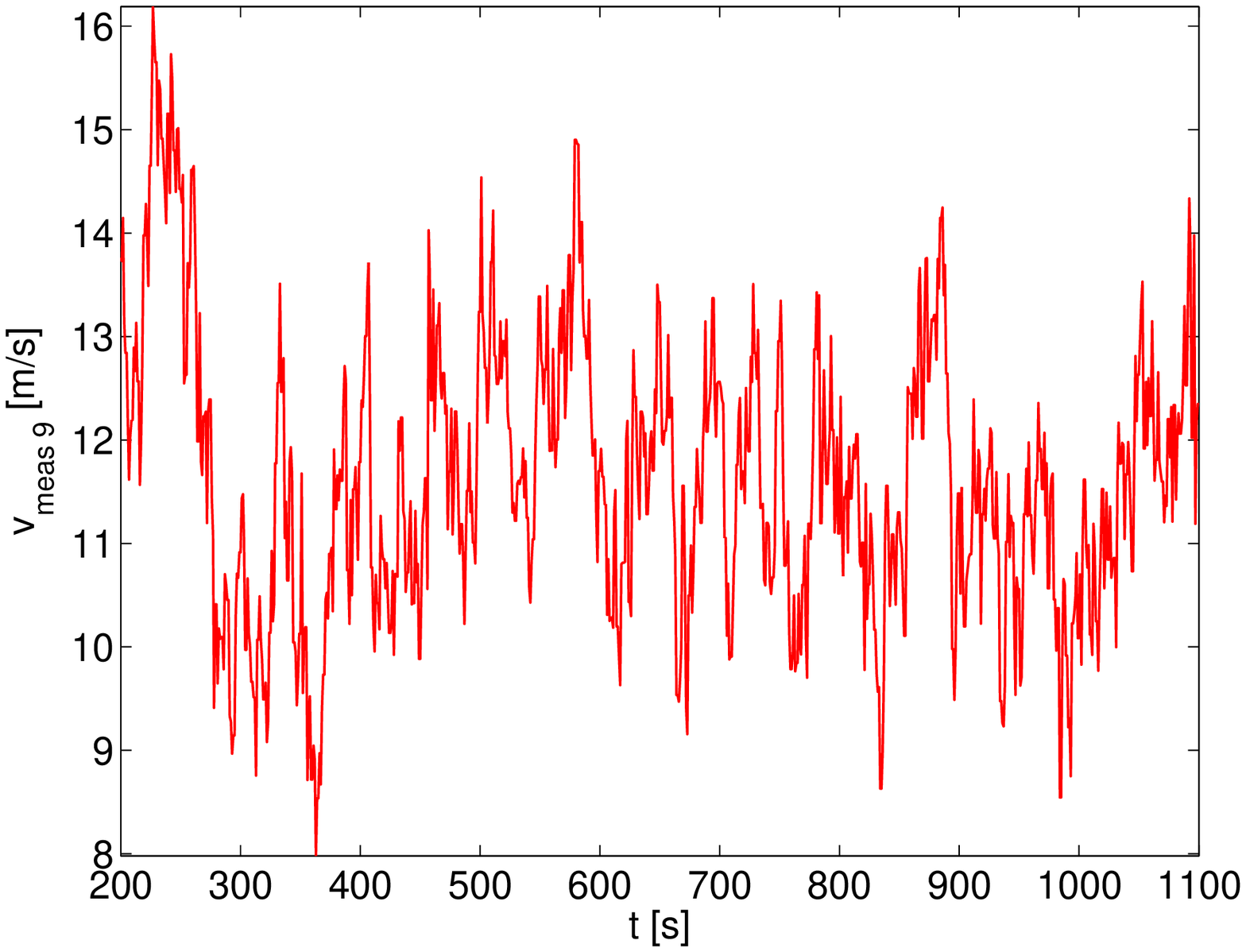}
                   \caption{Wind on WT $9$.}
                   \label{fig:WT9wind}
                 \end{subfigure}
                 \begin{subfigure}[!htb]{0.45\textwidth}
                   \centering
                   \includegraphics[scale=0.3]{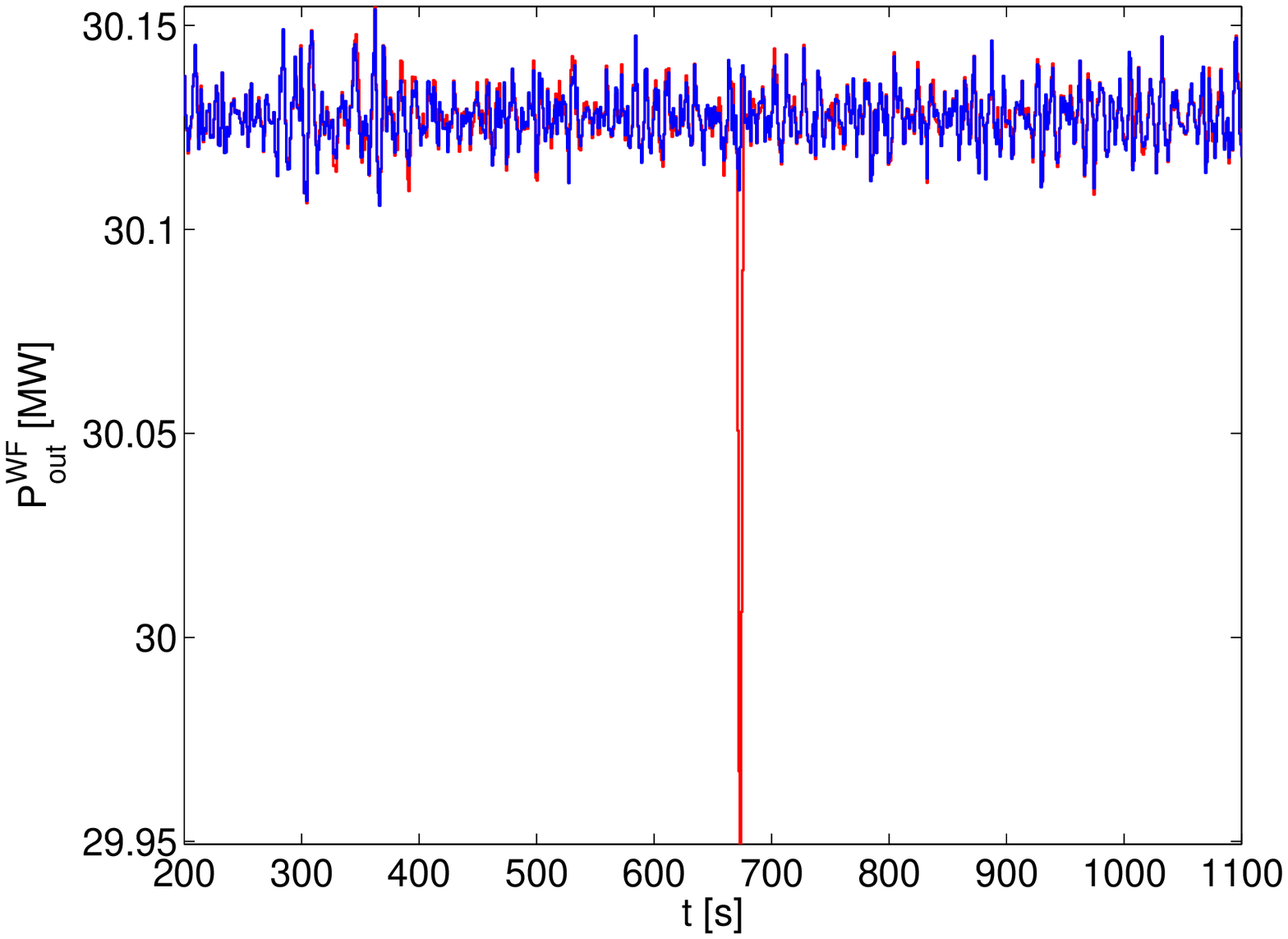}
                   \caption{Output WF power.}
                   \label{fig:WFpowerout}
                 \end{subfigure}
                 \caption{WF composed of $10$ WTs: comparison between scheduler controller, i.e. open-loop strategy (red) and DMPC, i.e. closed-loop strategy (blue). }
                 \label{fig:WF10results}
               \end{figure}

               In Figure \ref{fig:WF10results} we compare performance of scheduler (open-loop strategy) and DMPC (closed-loop strategy) in a single simulation. We note that using DMPC we achieve two aims: i) we guarantee that the WF produces the power demand required by the network operator (Figure \ref{fig:WFpowerout}) by removing the power drop of the open-loop strategy; ii) we reduce mechanical stress, by reducing variations in $M_s$ and $M_t$ (Figures \ref{fig:WT9mainshaft} and \ref{fig:WT9momentTower} for the $9$-th WT). We achieve our aims by changing the power demand set-points: in Figure \ref{fig:WT9powerdemand} (for the $9$-th WT) we highlight that instead of a constant power set-point, we allow to change $P_{dem,9}$ in a range of $0.1~[MW]$ that gives also good performance in the rate of change of the power demand set-point (usually $\abs{\frac{\partial P_{dem,i}}{\partial t}}\leq 0.1~[MW]$, see Figure \ref{fig:WT9gradientpowerdemand}).

          \subsection{Performance using different prediction horizons}
          	\label{sec:performancePredictionHorizon}
\begin{figure}[!htb]%
                 \centering
                 \includegraphics[scale=0.52]{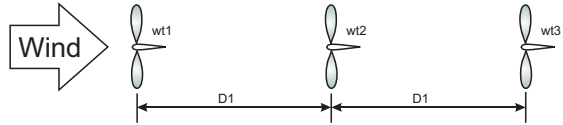}%
                 \caption{WF layout for example using $3$ WTs. $D1 = 400\;[m]$.}%
                 \label{fig:WF3}%
               \end{figure}

\begin{figure}[!htb]
                 \centering
                 \begin{subfigure}[!htb]{0.45\textwidth}
                   \centering
                   \includegraphics[scale=0.3]{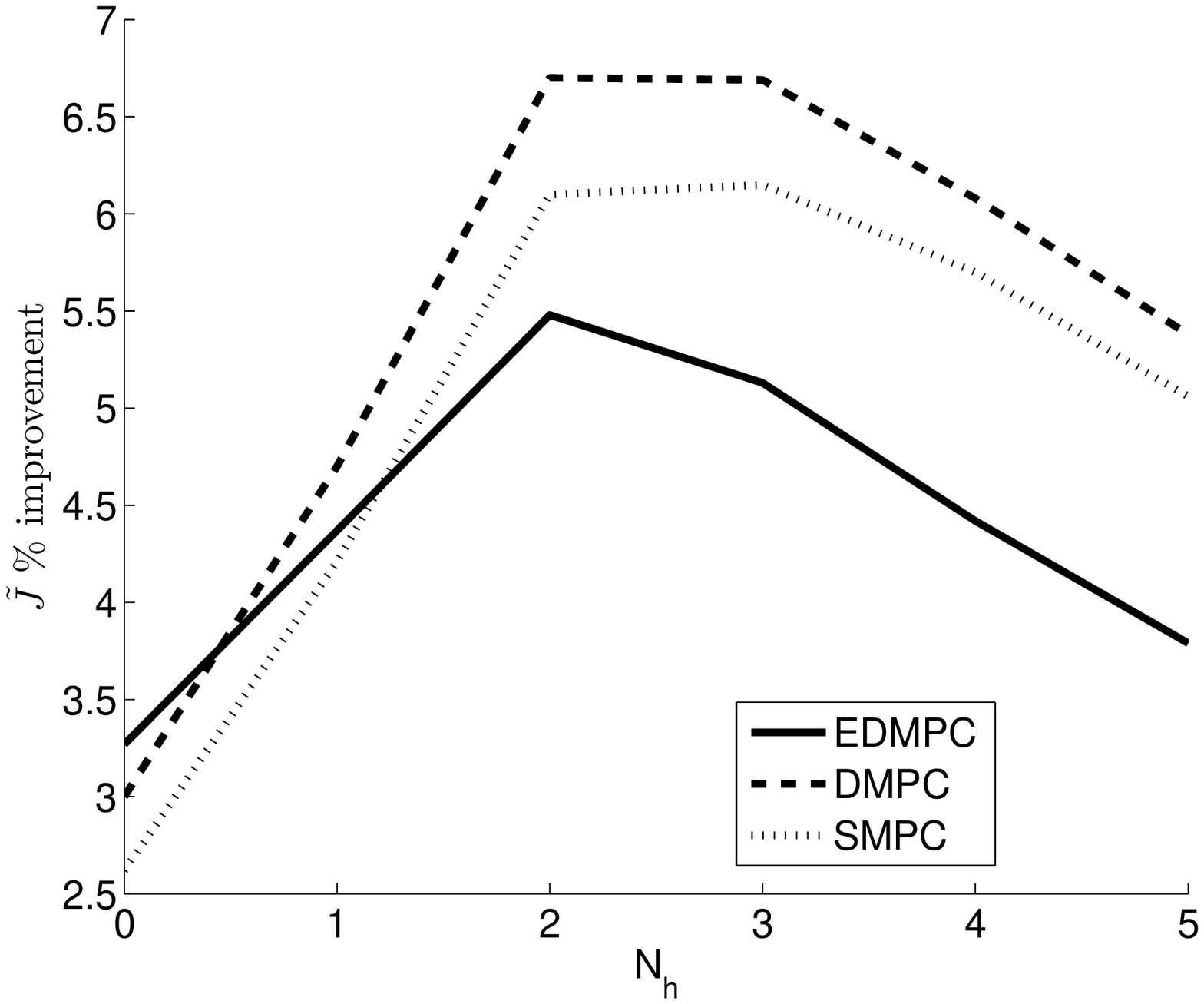}
                   \caption{Improvements in terms of $\tilde J$.}
                   \label{fig:WT3tildeJ}
                 \end{subfigure}
                 \begin{subfigure}[!htb]{0.45\textwidth}
                   \centering
                   \includegraphics[scale=0.3]{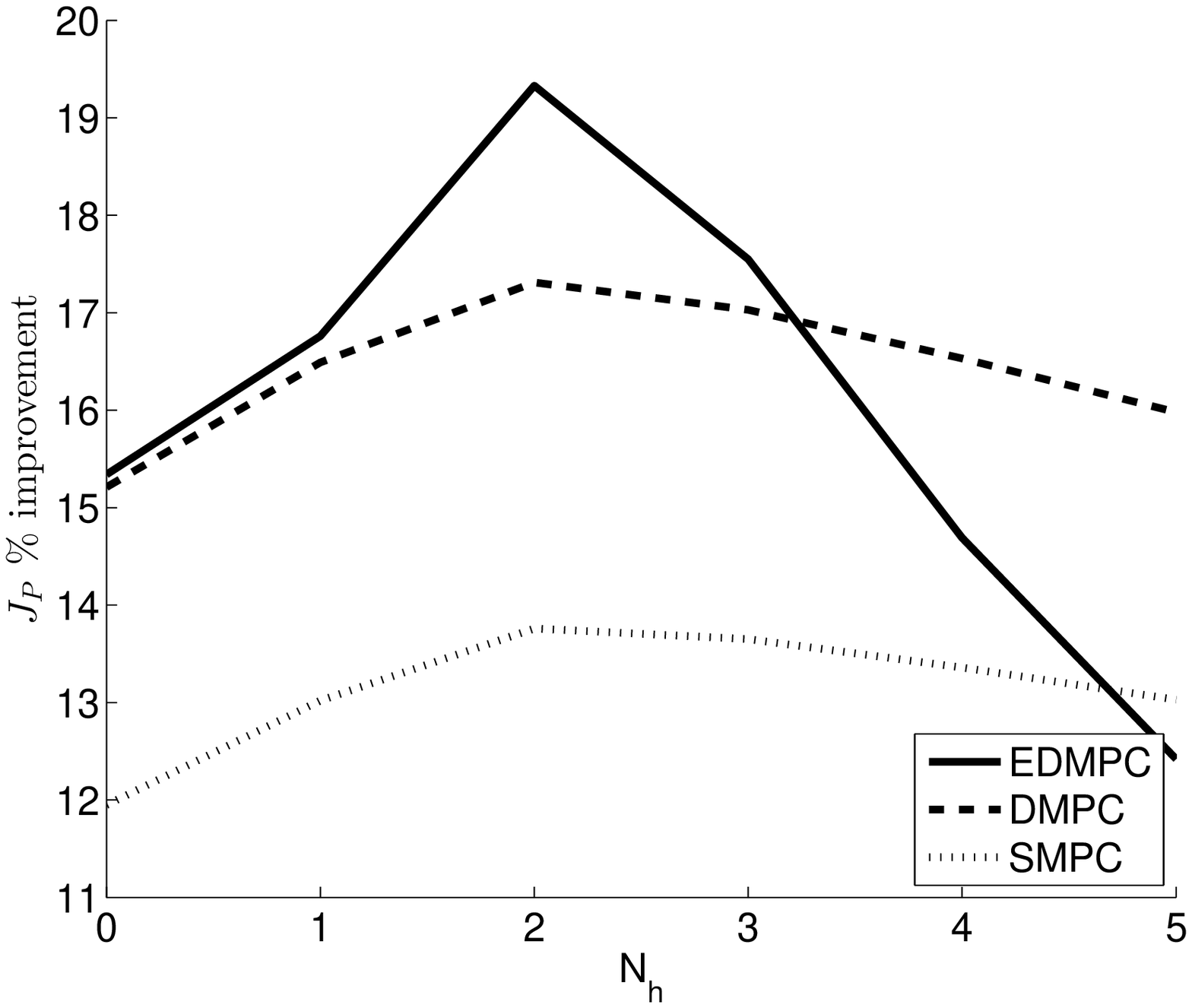}
                   \caption{Improvements in terms of $J_P$.}
                   \label{fig:WT3JP}
                 \end{subfigure}\\
                 \begin{subfigure}[!htb]{0.45\textwidth}
                   \centering
                   \includegraphics[scale=0.3]{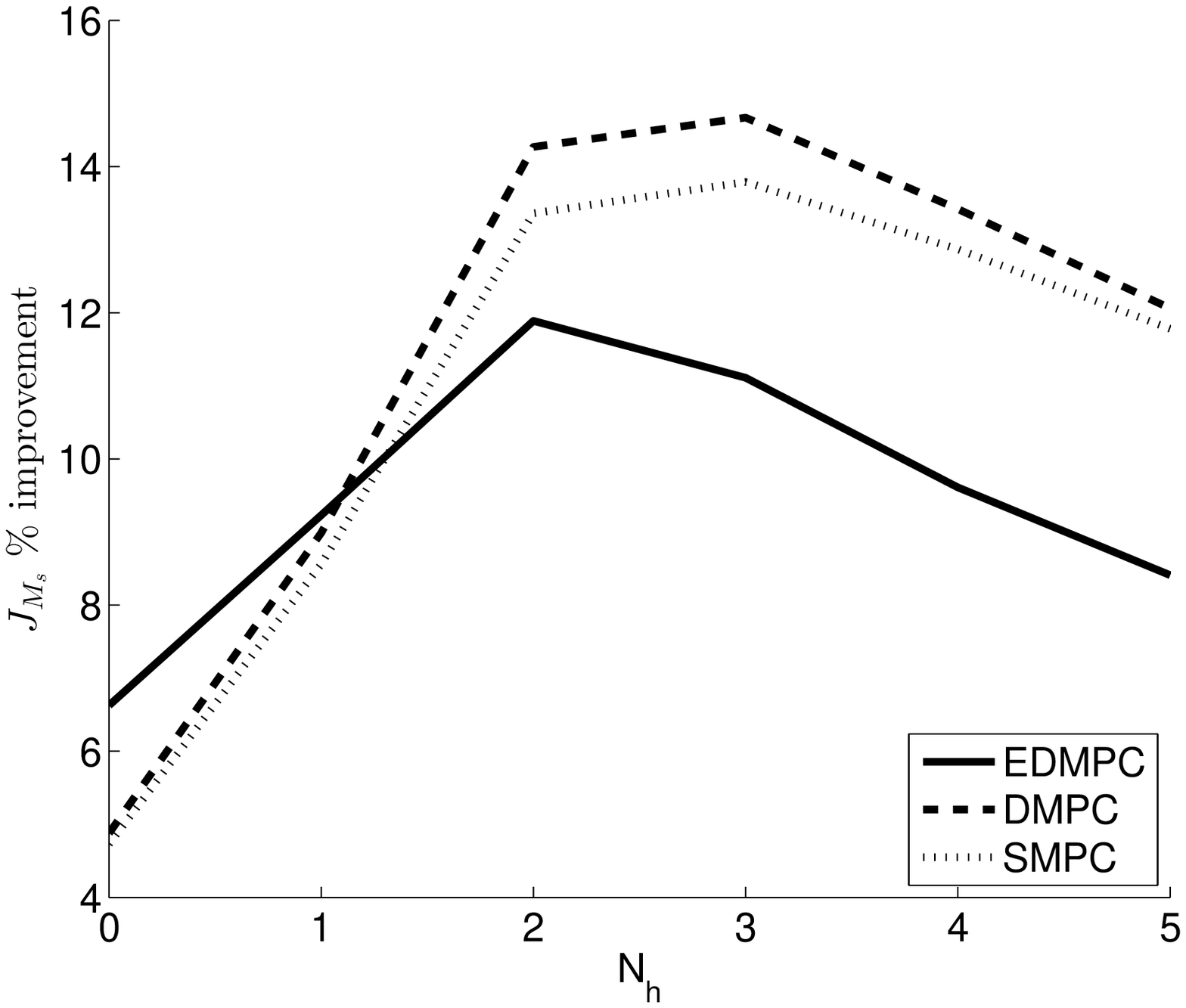}
                   \caption{Improvements in terms of $J_{M_s}$.}
                   \label{fig:WT3JMs}
                 \end{subfigure}
                 \begin{subfigure}[!htb]{0.45\textwidth}
                   \centering
                   \includegraphics[scale=0.3]{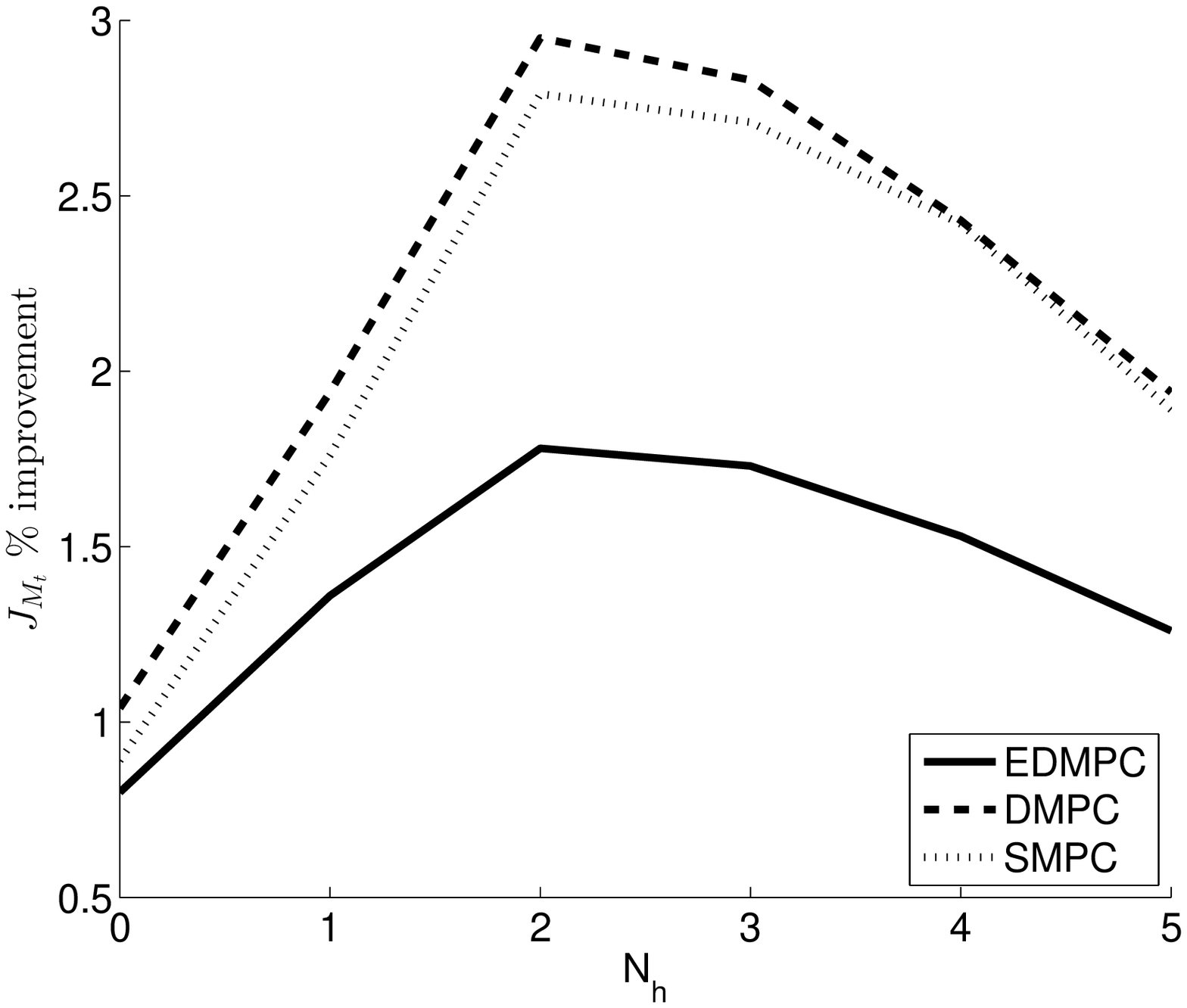}
                   \caption{Improvements in terms of $J_{M_t}$.}
                   \label{fig:WT3JMt}
                 \end{subfigure}
                 \caption{WF composed of $3$ WTs: comparison between MPC controllers using different prediction horizons $N_h$. Each plot has been obtained by averaging results obtained in $5$ simulations of $15$ minutes each. In all panels percentage of improvement with respect to the open-loop scheduler is shown.}
                 \label{fig:WF3resultsdifferentHorizon}
               \end{figure}

               In this section, we test the proposed MPC controllers in a WF composed of $3$ WTs arranged as shown in Figure \ref{fig:WF3}.
               The conditions of the WF and the regulator parameters are equal to those used in Section \ref{sec:WT10}. With this example we aim at studying performance for different prediction horizons. Results are shown in Figure \ref{fig:WF3resultsdifferentHorizon}. We note that for all MPC regulators maximum performance are achieved with prediction horizons $N_h=2$ and $N_h=3$. This means that, due to inaccurate wind predictions, performance decreases if the prediction horizon increases. Moreover we also note that designing an MPC controller with $N_h=0$ corresponds to dispatching the power demand based on the knowledge of current state and current wind measurements only. Hence, the prediction is one-step ahead only.

          \subsection{Performance without wind predictor}
               In this section, we test the proposed MPC controllers in a WF composed of $3$ WTs arranged as shown in Figure \ref{fig:WF3}. We use WF conditions and regulators parameters as in Section \ref{sec:performancePredictionHorizon}. Moreover we set $N_h=3$.
               \begin{figure}[!htb]
                 \centering
                 \begin{subfigure}[!htb]{0.45\textwidth}
                   \centering
                   \includegraphics[scale=0.3]{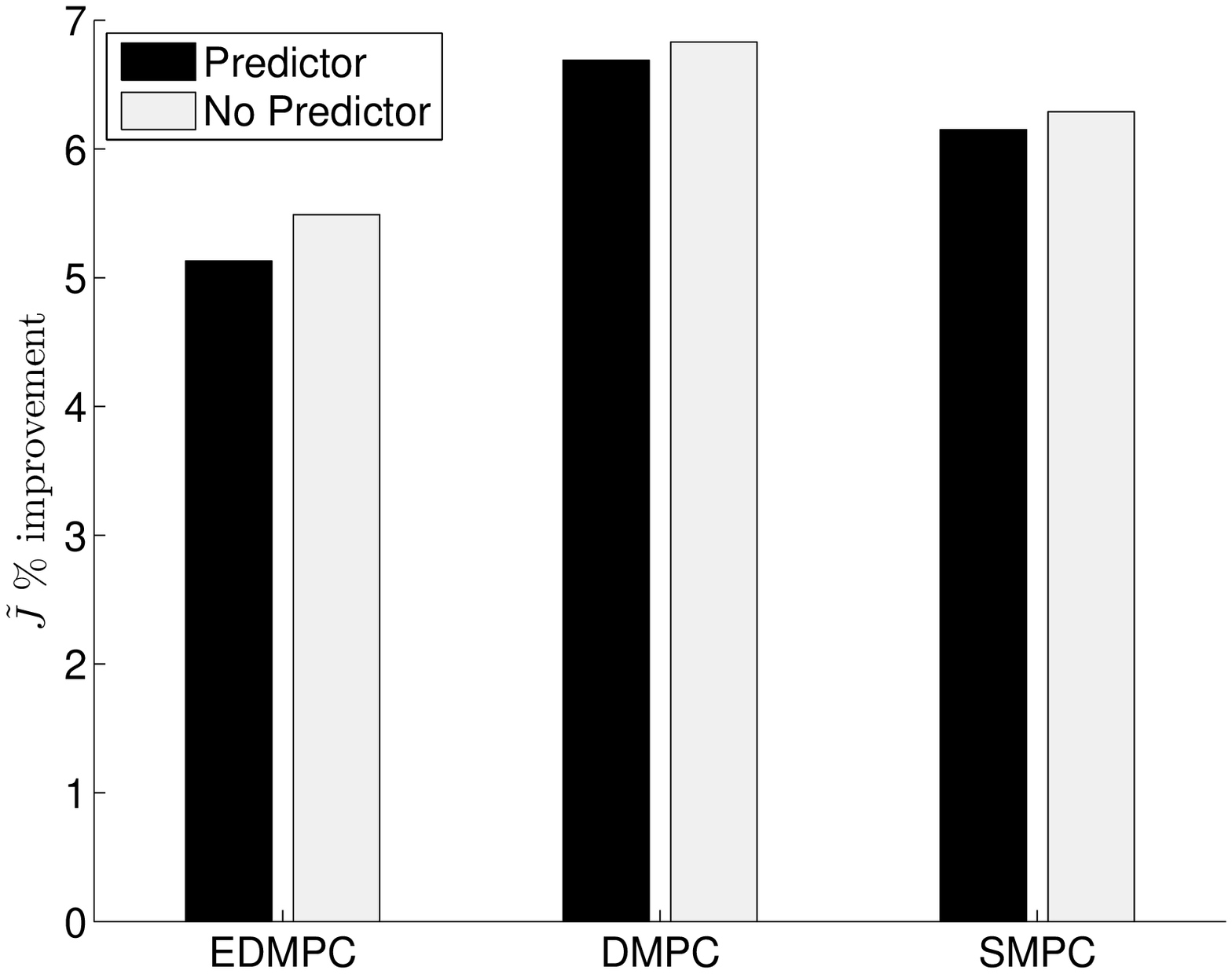}
                   \caption{Improvements in terms of $\tilde J$.}
                   \label{fig:WT3tildeJWP}
                 \end{subfigure}
                 \begin{subfigure}[!htb]{0.45\textwidth}
                   \centering
                   \includegraphics[scale=0.3]{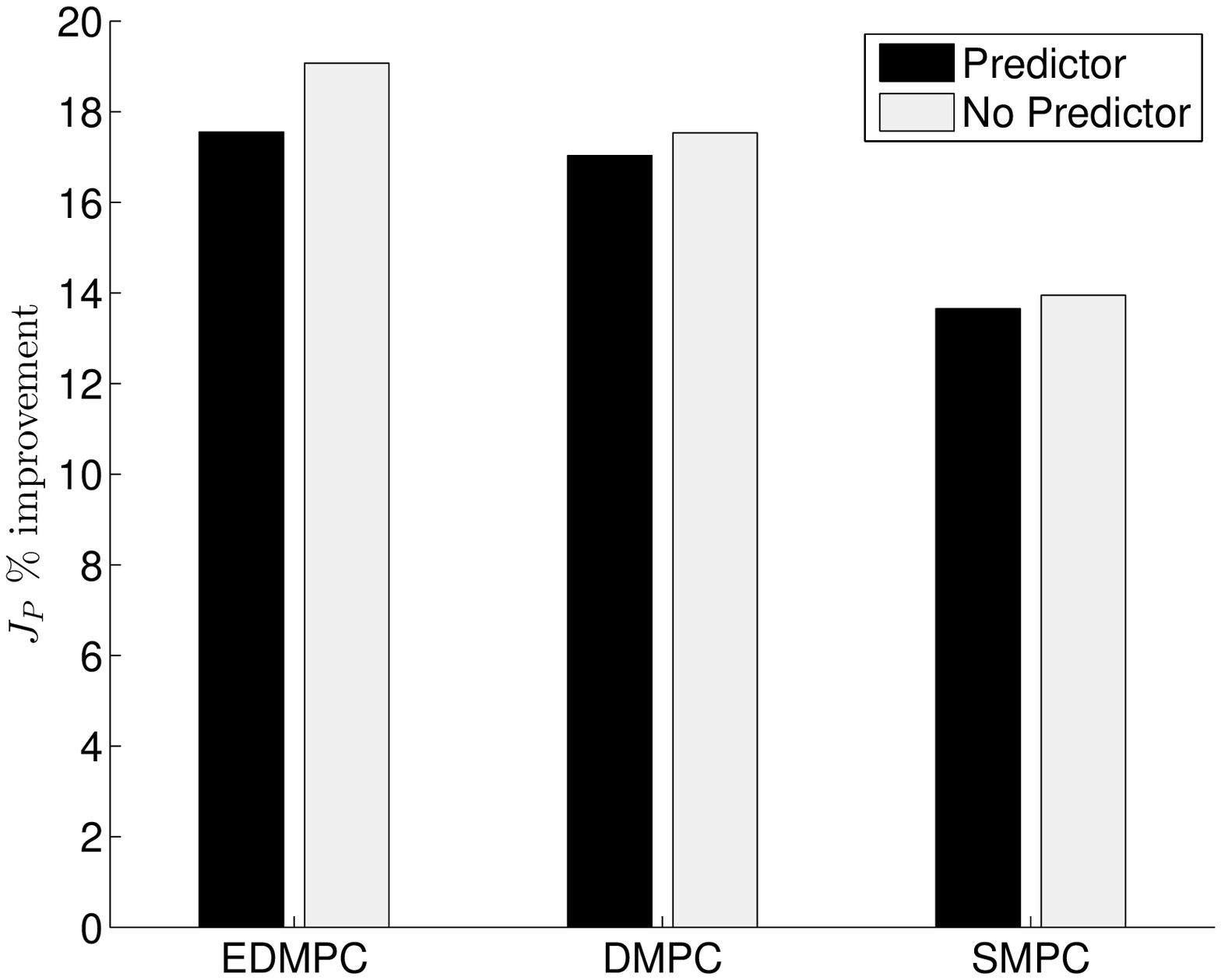}
                   \caption{Improvements in terms of $J_P$.}
                   \label{fig:WT3JPWP}
                 \end{subfigure}\\
                 \begin{subfigure}[!htb]{0.45\textwidth}
                   \centering
                   \includegraphics[scale=0.3]{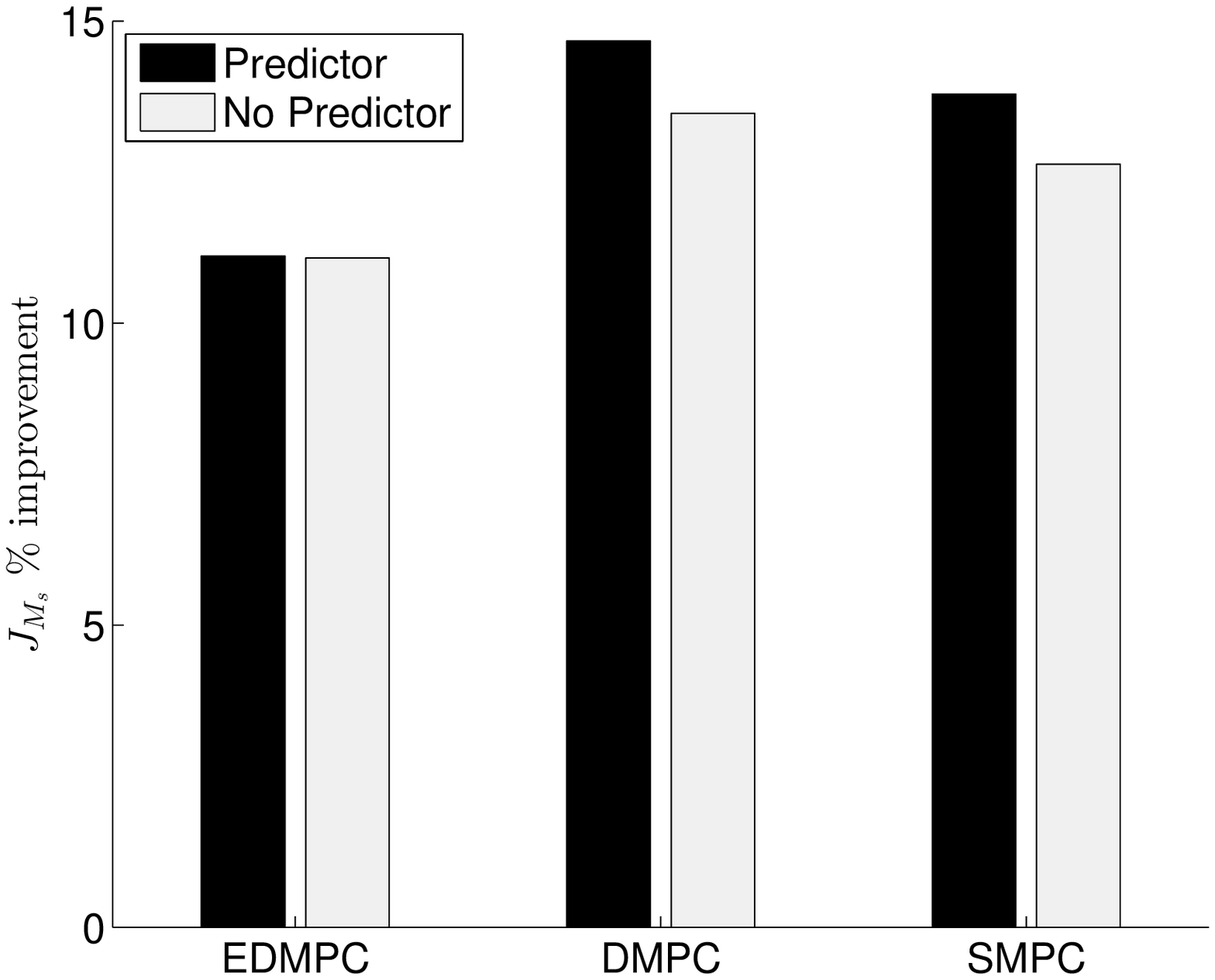}
                   \caption{Improvements in terms of $J_{M_s}$.}
                   \label{fig:WT3JMsWP}
                 \end{subfigure}
                 \begin{subfigure}[!htb]{0.45\textwidth}
                   \centering
                   \includegraphics[scale=0.3]{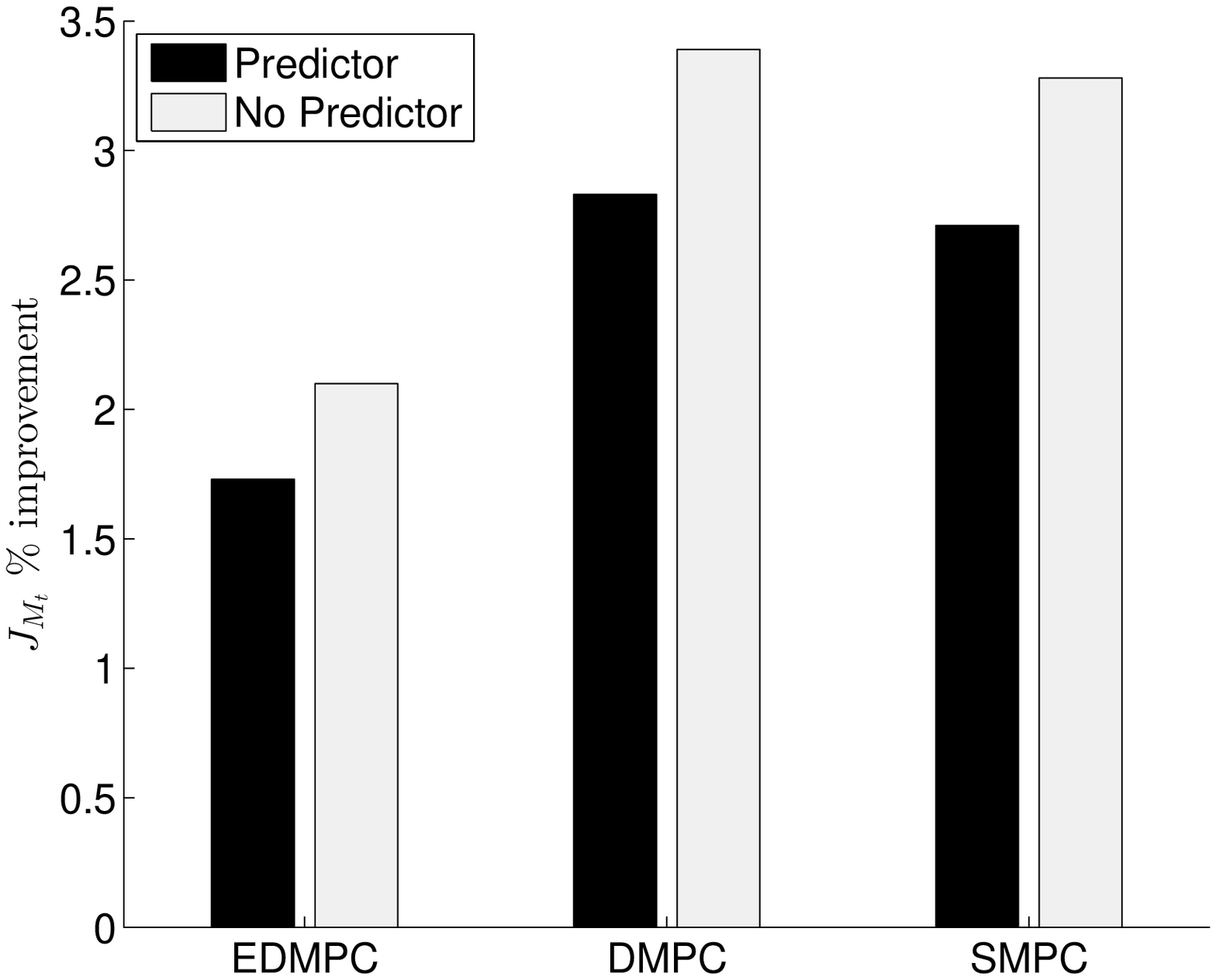}
                   \caption{Improvements in terms of $J_{M_t}$.}
                   \label{fig:WT3JMtWP}
                 \end{subfigure}
                 \caption{WF composed of $3$ WTs: comparison between MPC controllers using optimal wind one-step-ahead predictor (blue) and without using (red). Each barplot has been obtained by averaging results obtained in $5$ simulations of $15$ minutes each. In all panels, percentages of improvement with respect to the open-loop scheduler are shown.}
                 \label{fig:WF3resultsWP}
               \end{figure}
               In Figure \ref{fig:WF3resultsWP} we show performance with and without using the optimal wind one-step-ahead predictor. We note that for $\tJ$, $J_P$ and $J_{M_t}$ the use of the wind predictor decreases the performance. However $J_{M_s}$ increases, in particular using DMPC and SMPC. The reasons are the following: i) for the linearized output $M_{t,i}$ we do not consider any elastic model of the tower oscillations that depend on the wind acting on the tower (see \cite{Spudic2010a}); ii) the optimal wind predictor is designed locally for each WT and hence it does not account for wind interactions (see \cite{Johnson2009,Madjidian2011a}). These effects are more apparent if the wind turbulence increases. In future research we will also consider elastic model of tower oscillations and optimal wind predictors taking into account wind interactions among WTs. However, if our goal is to minimize main shaft fatigue only, the proposed wind predictors guarantee good performance.

          \subsection{Thanet Offshore Wind Farm}
\begin{figure}[!htb]%
                 \centering
                 \includegraphics[scale=0.45]{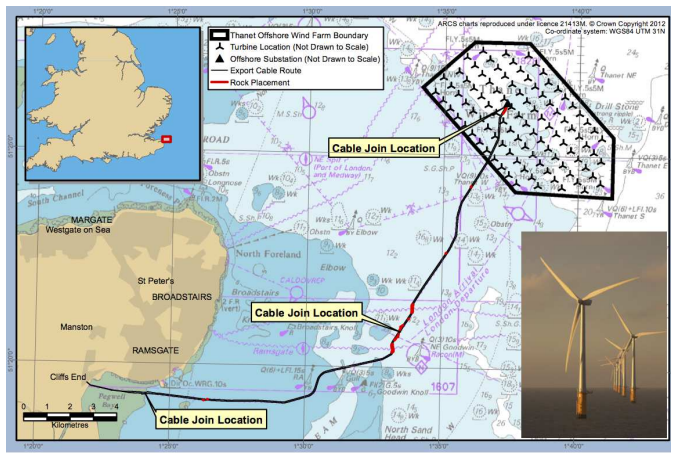}%
                 \caption{WF layout for the Thanet Offshore Wind Farm \cite{Vattenfall2014}.}
                 \label{fig:WF100}%
               \end{figure}

               In this last example, we consider the Thanet Offshore Wind Farm \cite{Vattenfall2014}, a WF composed of $100$ WTs arranged as shown in Figure \ref{fig:WF100}.
               For this example, we have imposed a WF power demand $P_{dem}^{WF} = 300\;[MW]$, equally distributed by the scheduler among the $100$ WTs, hence $P_{dem0_{i}} = 3\;[MW]$, $i=1:100$. Moreover, in the MPC cost function, we set $r_i=0.06$ for DMPC and $r_i=0.1$ for EDMPC and we require that $\abs{u_i(t)}\leq 0.1~[MW]$. The prediction horizon is $N_h=3$. The wind speed at the operating point is $v_{0} = 15\;\left[ \frac{m}{s} \right]$ and its turbulence is $T_I=0.1$. For this example, we were not able to use SMPC since, at every time instant, it requires the solution to a very large-scale SDP optimization problem.\\
               In Table \ref{tab:exampleWT100}, we summarize performance achieved by using different controllers. Compared with the open-loop controller and the SWFctrl, MPC controllers can diminish the mechanical stress of $13.07\%$ for $M_s$ and $2.13\%$ for $M_t$.
               \begin{table}[!htb]
                 \centering
                   \begin{tabular}{|c||c||c||c||c|}
                     \hline
                     & $\tJ$ & $J_P$ & $J_{M_s}$ &  $J_{M_t}$ \\
                     \hline
                     Scheduler only    &      3.7429    &   0.0024    &    1.2708    &    2.4698      \\
                     \hline
                     SWFctrl & -0.34\% &  0.01\%  &   -4.74\%   &  1.92\% \\
                     \hline
                     EDMPC & 4.07\% & 0.15\%  &   9.18\%   & 1.45\% \\
                     \hline
                     DMPC & \textbf{5.84\%} &  \textbf{0.22\%}  &   \textbf{13.07\%}   &  \textbf{2.13\%} \\
                     \hline
                   \end{tabular}
                 \caption{Controllers performance of the Thanet Offshore Wind Farm. Table entries have been obtained averaging $5$ simulations of $15$ minutes each. Top row: performances using open-loop scheduling. Other rows: percentage improvement with respect to the values in the first row. Best performances are in bold.}
                 \label{tab:exampleWT100}
               \end{table}

     \section{Conclusions}
          \label{sec:conclusions}
In this paper, we proposed MPC-based algorithms for dispatching a power demand for the whole WF among different WTs. The goal is to achieve minimization of the total mechanical stress. At the modeling level, we proposed to include in WT models a one-step ARMA predictor of the wind turbulence. We then demonstrated 
through simulations that this allows MPC dispatchers to achieve good performances in realistic scenarios. 
Future works will focus on increasing performances by improving the mechanical description of individual WTs as well as the model of the whole WF by accounting for interactions among WTs.
 \section{Acknowledgment}
       The authors are indebted with Dr. Vedrana Spudi\'c for insightful discussions as regards the linearized NREL model.

\appendix
\section{Derivation of the SMPC problem \eqref{prbl:mpcbase_stochastic}-\eqref{eq:LMIinputcons3}} \label{App:AppendixA}
First, we recall the following results that will be useful in the sequel.
          \begin{lem}[Schur Complement]
            \label{lem:schur}
            Let $Z= \begin{bmatrix} 
                A & B \\
                B^T & C \end{bmatrix}$ be a symmetric matrix partitioned into blocks $A,~B,~C$ where both $A$ and $C$ are symmetric and square. Assume that $A$ is positive semi-definite and $C$ is positive definite. Let $S= A - BC^{-1}B^{T}$ be the Schur Complement of $C$ in $Z$. Then, $Z\geq 0$ if and only if $S\geq 0$.
          \end{lem}

                   From \eqref{eq:xconstr_base_t}, the mean value state dynamics can be obtained by neglecting $w_k$, and it is given by \eqref{eq:nominalDyn}, i.e.
               \begin{equation}
                 \label{eq:nominalDyn1}
                 \bx_{k+1} = A \bx_{k} + \hB \bar\hu_{k},~\forall k= t+1:t+N_h,
               \end{equation}
               where $\bx_k = \Eset [x_{k}]$, $\bar{\hat{u}}_k = \Eset [\hat{u}_{k}]$. Defining the error variable $\delta_{k} = x_{k} - \bar{x}_{k}$ and assuming a control law of the form
               \begin{equation}
                 \label{eq:controlLaw}
                 \hat{u}_{k} = \bar{\hat{u}}_{k} + K_{k} \delta_{k} 
               \end{equation}
               where $K_{k} \in \Rset^{{N-1} \times n}$, one has that, for $\forall k= t+1:t+N_h$, $\delta_{k}$ is zero-mean Gaussian random variable with covariance matrix $X_{k}$ evolving as
               \begin{align}
                 \label{eq:dynCov}
                 X_{k+1} = \Eset\{\delta_{k+1}\delta_{k+1}^{T}\} = (A + \hat{B}K_{k})X_{k}(A + \hat{B}K_{k})^{T} + B_{d} \Sigma_{w} B_{d}^{T}.
               \end{align}
               Moreover, $x_{k} \sim \NN(\bar{x}_k,X_{k})$, for $\forall k= t+1:t+N_h+1$.
               \begin{rmk}
                 We note that since it is possible to measure online $\beta$,\ $\omega_{r}$,\ $\omega_{g}^{filt}$ (through the sensors placed on each WT) and the states of the optimal predictor, we can assert that the state of the system at time $t$ is measurable. Moreover, since we can also measure wind speed for each WT, we can affirm that $x_{t+1}$ is not affected by stochasticity (see also \eqref{eq:xconstr_base_0}). Therefore, \eqref{eq:dynCov} is initialized with \eqref{eq:x0x1}, i.e.
                 \begin{equation}
                   X_{t} = X_{t+1} = 0 \label{eq:x0x1_1}
                 \end{equation}
                 and, always according to \eqref{eq:dynCov}, we also have \eqref{eq:x2}, i.e.
                 \begin{equation}
                   X_{t+2}= B_{d} \Sigma_{w} B_{d}^{T}.\label{eq:x2_1}
                 \end{equation}
               \end{rmk}
               We highlight that, since \eqref{eq:dynCov} depends both from variables $X_k$ and $K_k$, the dynamics of the covariance matrix is nonlinear. However, by relaxing constraint \eqref{eq:dynCov} from equality to inequality constraint and using Lemma \ref{lem:schur}, we can rewrite \eqref{eq:dynCov} as \eqref{eq:LMIvarianceDyn}, i.e.
               \begin{equation}
                 \label{eq:LMIvarianceDyn_1}
                 \matr{
                   X_{k+1} & AX_{k}+\hat{B}G_{k} & B_{d}\Sigma_{w}\\
                   (\ast) & X_{k} & 0 \\
                   (\ast) & (\ast) & \Sigma_{w}
                 } \geq 0,\quad k=t+3:t+N_h
               \end{equation}
               where $G_k=K_kX_k$ and $(\ast)$ denotes the matrix transpose of the corresponding block in the upper triangular part. In other words, if there are $X_k$, $X_{k+1}$ and $G_k$ verifying \eqref{eq:LMIvarianceDyn_1}, then one has $X_{k+1}\geq \Eset[\delta_{k+1}\delta_{k+1}^T]$.\\               
               Next, using \eqref{eq:yconstr_base_0} and \eqref{eq:yconstr_base_t}, we rewrite cost function \eqref{eq:costMPC_base} as
               \begin{equation}
                 \label{eq:costMPC_base_stochastic}
                 \Eset\left[\norme{\matr{x_{t}\\\hu_t\\d_t}}{M_0}^2\right] + \sum_{k=t+1}^{t+N_h} \Eset\left[\norme{\matr{x_{k}\\\hu_k\\w_k}}{M}^2\right]
               \end{equation}
               where 
               $$
               M_0 = \matr{ C_0^TQC_0 & C_0^TQ\hD & C_0^TQD_d \\  (\ast) & T^TRT+\hD^TQD & \hD^TQD_d \\  (\ast) &  (\ast) & D_d^TQD_d},
               \quad
               M = \matr{ C^TQC & C^TQ\hD & C^TQD_d \\  (\ast) & T^TRT+\hD^TQD & \hD^TQD_d \\  (\ast) &  (\ast) & D_d^TQD_d}.
               $$
               Our next aim is to remove averages in the cost function \eqref{eq:costMPC_base_stochastic}. For this purpose, we proceed as described in \cite{Magni2009a}. We recall the following properties.
               \begin{equation}
                 \begin{aligned}
                   \Eset \left[ X^{T} F X \right] = \Eset \left[ tr \left( F X X^{T} \right) \right]= tr \left( \Eset \left[F X X^{T} \right] \right)= tr \left(F\Eset \left[X X^{T} \right] \right) \label{eq:Exvalproof}
                 \end{aligned}
               \end{equation}
               \begin{equation}
                 \Eset \left[ X X^{T} \right] =  var \left[ X \right] + \Eset \left[ X \right] \Eset \left[ X^{T} \right] \label{eq:varproof}
               \end{equation}
               where $X$ is a random vector and $F$ is a square positive semi-definite matrix.\\
               Applying \eqref{eq:Exvalproof} to \eqref{eq:costMPC_base_stochastic}, we obtain
               \begin{equation*}
                 tr\left(M_0 \Eset\left[ \matr{x_{t}\\\hu_t\\d_t} \matr{x_{t}\\\hu_t\\d_t}^T \right]\right) + \sum\limits_{k=t+1}^{N_h} tr\left(M \underbrace{\Eset\left[ \matr{x_{k}\\\hu_k\\w_k} \matr{x_{k}\\\hu_k\\w_k}^T \right]}_{(\bullet)}\right).
               \end{equation*}               
               Applying \eqref{eq:varproof} to the highlighted part $(\bullet)$ we have
               \begin{equation*}
                 \begin{split}
                   &\Eset\left[ \matr{x_{k}\\\hu_k\\w_k} \matr{x_{k}\\\hu_k\\w_k}^T \right] \\
                   &= \Eset\left[ \matr{x_{k}\\\hu_k\\w_k}\right]\Eset\left[ \matr{x_{k}^T & \hu_k^T & w_k^T } \right] + var\left[\matr{x_{k}\\\hu_k\\w_k}\right]\\
                   &= \matr{\Eset[x_{k}]\\\Eset[\hu_k]\\\Eset[w_k]} \matr{\Eset[x_{k}]\\\Eset[\hu_k]\\\Eset[w_k]}^T + \matr{ var[x_k] & cov[x_k,\hu_k] & cov[x_k,w_k]  \\ cov[\hu_k,x_k] & var[\hu_k] &cov[\hu_k,w_k] \\ cov[w_k,x_k] & cov[w_k,\hu_k] & var[w_k] } \\ 
                   &= \matr{\Eset[x_{k}]\\\Eset[\hu_k]\\0} \matr{\Eset[x_{k}]\\\Eset[\hu_k]\\0}^T + \matr{ var[x_k] & var[x_k]K_k^T & 0  \\ K_kvar[x_k] & K_kvar[x_k]K_k^T & 0 \\ 0 & 0 & \Sigma_w } \\                  
                   &= \matr{\Eset[x_{k}]\\\Eset[\hu_k]\\0} \matr{\Eset[x_{k}]\\\Eset[\hu_k]\\0}^T + \matr{ var[x_k] & 0 \\ K_kvar[x_k] & 0 \\ 0 & \Iset }\matr{ var[x_k] & 0 \\ 0 & \Sigma_w^{-1}}^{-1}\matr{ var[x_k]^T & var[x_k]K_k^T & 0 \\ 0 & 0 & \Iset}.
                 \end{split}
               \end{equation*}
               Replacing $\Eset \left[ x_{k} \right]$,\ $\Eset \left[ \hat{u}_k \right]$,\ $var \left[ x_{k} \right]$ and \ $K_{k} var \left[ x_{k} \right]$ respectively with $\bar{x}_{k}$,\ $\bar{\hat{u}}_{k}$,\ $X_{k}$ and $G_{k}$, we obtain
               \begin{equation}
                 \label{eq:ExvalforPk}
                 \Eset\left[ \matr{x_{k}\\\hu_k\\w_k} \matr{x_{k}\\\hu_k\\w_k}^T \right] = \matr{\bx_k\\\bar\hu_k\\0} \matr{\bx_k\\\bar\hu_k\\0}^T + \matr{ X_k & 0 \\ G_k & 0 \\ 0 & \Iset }\matr{ X_k & 0 \\ 0 & \Sigma_w^{-1}}^{-1}\matr{ X_k^T & G_k^T & 0 \\ 0 & 0 & \Iset}.
               \end{equation}
               This relation is valid $\forall k = t+2,\ldots,N_h$, since, for these values of $k$,  matrices $X_{k}$ are positive-definite and therefore invertible. For the time instants $j = t,t+1$, we have
               \begin{equation}
                 \label{eq:Pzero}
                 \Eset\left[ \matr{x_{j}\\\hu_j\\w_j} \matr{x_j\\\hu_j\\w_j}^T \right] = \matr{\bx_j\\\bar\hu_j\\0} \matr{\bx_j\\\bar\hu_j\\0}^T.
               \end{equation}
               Let us now define
               \begin{equation*}
                 P_{k}= \Eset\left[ \matr{x_{k}\\\hu_k\\w_k} \matr{x_{k}\\\hu_k\\w_k}^T \right]
               \end{equation*}
               Then, relaxing the equality constraint \eqref{eq:ExvalforPk}, we obtain
               \begin{equation}
                 \label{eq:ExValforPkIneq}
                 P_{k} \geq 
                 \matr{\bx_k\\\bar\hu_k\\0} \matr{\bx_k\\\bar\hu_k\\0}^T + \matr{ X_k & 0 \\ G_k & 0 \\ 0 & \Iset }\matr{ X_k & 0 \\ 0 & \Sigma_w^{-1}}^{-1}\matr{ X_k^T & G_k^T & 0 \\ 0 & 0 & \Iset}.
               \end{equation}
               Applying Lemma \ref{lem:schur}, we rewrite \eqref{eq:ExValforPkIneq} as the LMI \eqref{eq:Pk}, i.e.
               \begin{align}
                 \label{eq:Pk_1}
                 \begin{bmatrix}
                   P_{k} & \matr{ X_{k} & 0 \\ G_{k} & 0 \\ 0 & \Iset } & \matr{\bx_k\\\bar\hu_k\\0} \\
                   (\ast) & \matr{ X_{k} & 0 \\ 0 & \Sigma_w^{-1} } & 0 \\
                   (\ast) & (\ast) & 1
                 \end{bmatrix},&\geq 0 &\forall k = t+2, \ldots, N_h.
               \end{align}
              Similarly, for \eqref{eq:Pzero}, introducing $P_{j}$, $j=t,t+1$, we obtain the  LMI \eqref{eq:Pj}, i.e.
               \begin{align}
                 \label{eq:Pj_1}
                 \begin{bmatrix}
                   P_{j} & \begin{bmatrix} \bar{x}_{j} \\ \bar{\hat{u}}_{j} \\0 \end{bmatrix} \\
                   (\ast)& 1
                 \end{bmatrix}&\geq 0.
               \end{align}
As a whole, an upper bound to the cost function in \eqref{eq:costMPC_base} is provided by the cost in \eqref{prbl:mpcbase_stochastic}.

               Our last aim is to account for probabilistic input constraints \eqref{eq:uconstrdis_base} using the procedure proposed in \cite{Magni2009a}. In the following, for simplicity of notation, we neglect the index $s$ and the time $k$ appearing in \eqref{eq:uconstrdis_base}. Suppose we want to impose
               \begin{equation}
                 \label{eq:LinCst}
                 \PP(\hat{c}^T \hat{u} \geq u^{max}) \leq \tp,
               \end{equation}
               Note that
               \begin{equation*}
\begin{aligned}
                 \PP(\hat{c}^T \hat{u} \geq u^{max}) &= \PP\left( \frac{\hat{c}^T \hat{u} - \Eset[\hat{c}^T \hat{u}]}{\sqrt{\hc^{T} var[\hat{u}] \hc}} \geq \frac{\hat{u}^{max} - \Eset[\hat{c}^T \hat{u}]}{\sqrt{\hc^{T} var[\hat{u}] \hc}} \right)=\\
&=1 - \PP\left( \frac{\hat{c}^T \hat{u} - \Eset[\hat{c}^T \hat{u}]}{\sqrt{\hc^{T} var[\hat{u}] \hc}} \leq \frac{\hat{u}^{max} - \Eset[\hat{c}^T \hat{u}]}{\sqrt{\hc^{T} var[\hat{u}] \hc}} \right)
                 \label{eq:lincostodet}
  \end{aligned}
               \end{equation*}
               Since $\hu$ is given by \eqref{eq:controlLaw}, where $\delta_k$ is Gaussian, one has that the
random variable $\frac{\hat{c}^T \hat{u} - \Eset[\hat{c}^T \hat{u}]}{\sqrt{\hc^{T} var[\hat{u}]\hc}}$ is distributed as $\NN(0,1)$. Hence we can write
               \begin{equation*}
                 \begin{aligned}
                   \PP(\hat{c}^T \hat{u} \geq u^{max})=1 - \mathcal{G} \left(\frac{\hat{u}^{max} - \Eset[\hat{c}^T \hat{u}]}{\sqrt{\hc^{T} var[\hat{u}] \hc}} \right).
                 \end{aligned}
               \end{equation*}
where $ \mathcal{G}(x)$ is the standard Gaussian probability distribution. Therefore, we can rewrite \eqref{eq:LinCst} as
               \begin{equation*}
                 \mathcal{G} \left(\frac{\hat{u}^{max} - \Eset[\hat{c}^T \hat{u}]}{\sqrt{\hc^{T} var[\hat{u}] \hc}} \right) \geq 1 - \tp.
               \end{equation*}
               Note that $\mathcal{G}$ is strictly monotone and invertible. Hence $\GG^{-1}$ is strictly monotone as well. Therefore we can state
               \begin{equation*}
                 \frac{\hat{u}^{max} - \Eset[\hat{c}^T \hat{u}]}{\sqrt{\hc^{T} var[\hat{u}] \hc}}  \geq \GG^{-1} \left( 1 - \tp \right).
               \end{equation*}
               In conclusion we obtained the deterministic constraint
               \begin{equation*}
                 \Eset[\hat{c}^T \hat{u}]  \leq \hat{u}^{max} - \left(\sqrt{\hc^{T} var[\hat{u}] \hc} \right) \GG^{-1} \left( 1 - \tp \right).
               \end{equation*}
               This new deterministic constraint, which involves the expected value of the random variable $\hat{c}^T\hat{u}$, can be rewritten using the Gauss's error function $erf(x)$, which verifies
               \begin{equation*}
                 \mathcal{G} (x) = \frac{1}{2} (1+erf(\frac{x}{\sqrt{2}}))
                 \label{eq:Gauss}
               \end{equation*}
               as
               \begin{equation}\label{eq:ecuGauss}
                 \Eset[\hat{c}^T \hat{u}]  \leq \hat{u}^{max} - \left(\sqrt{\hc^{T} var[\hat{u}] \hc} \right) \sqrt{2} erf^{-1} \left( 1 - 2\tp \right).
               \end{equation}
Constraint \eqref{eq:ecuGauss} is equivalent to the existence of $\Eset[\hat{u}]$, $var[\hat{u}]$ and $\psi>0$ such that, simultaneously,
\begin{align}
 \hat{c}^T var[\hat{u}] \hat{c} &\leq \psi^{2} \frac{1}{2} \left( \frac{1}{erf^{-1}(1-2\tp)} \right)^{2} 
                 \label{eq:boundvar} \\
\hat{c}^T\Eset \left[ \hat{u} \right] &\leq  u^{max} -\sqrt{\psi^2}
                 \label{eq:ecuGaussGamma}
\end{align}
Considering $\psi^2$ as optimization variable, instead of $\psi$, we note that it enters \eqref{eq:ecuGaussGamma} in a nonlinear way. In order to obtain an affine constraint, we linearize $\sqrt{\psi^2}$ about $\psi^{2} \approx \frac{(u^{max})^{2}}{4}$ (observe that $0\leq\psi\leq u^{max}$ and hence the linearization point lies in the middle of the interval). We get
\begin{equation}
                 \sqrt{\psi^{2}} \approx \frac{u^{max}}{4} + \frac{\psi^2}{u^{max}}.
                 \label{eq:rootBeta}
               \end{equation}
Summarizing, constraint \eqref{eq:ecuGauss} is replaced with \eqref{eq:boundvar} and 
\begin{align*}
                \hat{c}^T \Eset \left[ \hat{u} \right] &\leq \frac{3}{4} u^{max} - \frac{\psi^{2}}{u^{max}}.
                 \label{eq:ctvaruc}
               \end{align*}
               We highlight that since from \eqref{eq:controlLaw} the control input $\hat{u}_{k}$ depends on the Gaussian error $\delta_{k}$, we have that $\hat{u}_{k} \sim WGN(\bar{\hat{u}}_{k},K_{k}X_{k})$, for all $k = t:t+N_h$, and therefore the control law variance must be assumed as optimization variable. The expected value $\Eset[\hat{u}]$ is $\bar{\hat{u}}$, while the related variance depends on $var[x_{k}]$ and $K_{k}$. In fact we have
               \begin{equation*}
                 \begin{aligned}
                   var[\hat{u}_k] &= var\left[ \bar{\hat{u}}_{k} + K_{k} \delta_{k} \right] = var\left[ \bar{\hat{u}}_{k} + K_{k} (x_{k}- \Eset[x_{k}] ) \right]\\
                   &= var\left[ \bar{\hat{u}}_{k} + K_{k} (x_{k}- \bar{x}_{k} ) \right] = var\left[ K_{k} (x_{k}- \bar{x}_{k} ) \right]\\
                   &= K_{k} var\left[ (x_{k}- \bar{x}_{k} ) \right] K_{k}^{T} = K_{k} X_{k} K_{k}^{T}
                 \end{aligned}
                 \label{eq:varu}
               \end{equation*}
               and substituting $G_k= K_{k} X_{k}$ we obtain
               \begin{equation}
                 \begin{aligned}
                   var[\hat{u}_k] = K_{k} X_{k} X_{k}^{-1} X_{k} K_{k}^{T} = G_{k} X_{k}^{-1} G_{k}^{T}.
                 \end{aligned}
                 \label{eq:varuGk}
               \end{equation}
               Now, in order to obtain an LMI constraint, we relax \eqref{eq:varuGk} as
               \begin{equation*}
                 var[\hat{u}_k] - G_{k} X_{k}^{-1} G_{k}^{T} \geq 0
                 \label{eq:varuSchur1}
               \end{equation*}
               and applying Lemma \ref{lem:schur}, we have the constraint
               \begin{equation*}
                 \begin{bmatrix}
                   var[\hat{u}_k] & G_{k}\\
                   G_{k}^{T} & X_{k}         
                 \end{bmatrix} \geq 0,\qquad k = t:t+N_h.
                 \label{eq:uconstrLMI}
               \end{equation*}
               Concluding, to manage probabilistic linear input constraints, we replace \eqref{eq:uconstrdis_base} with 
\eqref{eq:LMIinputcons1}-\eqref{eq:LMIinputcons3}, i.e.
               \begin{align}                 
                 &\label{eq:LMIinputcons1_1} \begin{bmatrix} \hat{U}_{k} & G_{k}\\
                   G_{k}^{T} & X_{k} \end{bmatrix} \geq 0,\quad k = t:t+N_h\\
                 &\label{eq:LMIinputcons2_1} \hat{c}_{s}^T \bar{\hat{u}}_{k} \leq \frac{3}{4} u_{s}^{max} - \frac{\theta_{ks}}{u_{s}^{max}},\quad k = t:t+N_h \text{ and } s = 1:S \\
                 &\label{eq:LMIinputcons3_1} \hat{c}_{s}^{T} \hat{U}_{k} \hat{c}_{s} \leq \theta_{ks} \frac{1}{2} \left( \frac{1}{erf^{-1}(1 - 2\tp)} \right)^{2},\quad k = t:t+N_h \text{ and } s = 1:S
               \end{align}
               where $\hat{U}_{k} = var[\hat{u}_{k}]$, $\Eset \left[ \hat{c}_{s} \hat{u}_{s} \right] = \hat{c}_{s} \bar{\hat{u}}_{s}$ and $\theta_{ks} = \psi_{ks}^{2}$. \\               
Summarizing all above results, the optimization problem that must be solved online at each time instant $t$ is  \eqref{prbl:mpcbase_stochastic}-\eqref{eq:LMIinputcons3}.

     \bibliographystyle{IEEEtran}
     \bibliography{IEEEabrv,MPC_Wind_Farms-report}

\end{document}